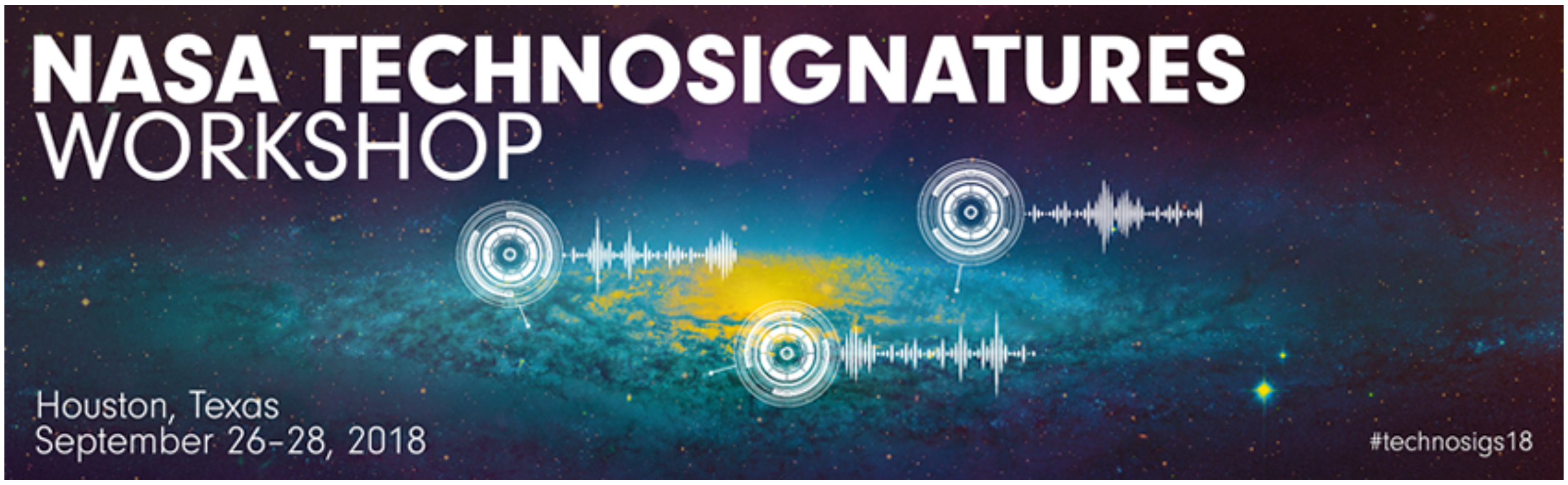

# NASA AND THE SEARCH FOR TECHNOSIGNATURES

## A Report from the NASA Technosignatures Workshop

NOVEMBER 28, 2018

# CONTENTS

# 1 INTRODUCTION

Jason Wright

On September 26–28, 2018, NASA hosted the Technosignatures Workshop at the Lunar and Planetary Institute in Houston, Texas, to learn more about the current field and state of the art for technosignatures searches and what role NASA could possibly play in the future in these searches.

The workshop had four main goals.

> 1. Define the current state of the technosignature field. What experiments have occurred? What is the state-of-the-art for technosignature detection? What limits do we currently have on technosignatures?
>
> 2. Understand the advances coming near-term in the technosignature field. What assets are in place that can be applied to the search for technosignatures? What planned and funded projects will advance the state-of-the-art in future years, and what is the nature of that advancement?
>
> 3. Understand the future potential of the technosignature field. What new surveys, new instruments, technology development, new data-mining algorithms, new theory and modeling, etc., would be important for future advances in the field?
>
> 4. What role can NASA partnerships with the private sector and philanthropic organizations play in advancing our understanding of the technosignatures field?

This report, edited by Jason Wright (PSU) and Dawn Gelino (NExScI), summarizes the material presented at the workshop, as well as additional inputs from the participants, and is organized as follows:

**Section 1, Introduction**, led by Jason Wright, explains the scope and purpose of this document, provides general background about the search for technosignatures, and gives context for what follows.

**Section 2, Current Upper Limits on Technosignatures,** led by Andrew Siemion, Emilio Enriquez, Shubham Kanodia, and Jason Wright, addresses components of the first goal of the workshop, including: *What experiments have occurred? What limits do we currently have on technosignatures?*

**Section 3, State of the Art of Searches for Technosignatures,** led by Andrew Siemion, Emilio Enriquez, Shubham Kanodia, and Jason Wright, addresses components of the first goal of the workshop, including: *Define the current state of the technosignature field. What is the state-of-the-art for technosignature detection?*

**Section 4, Near-term Searches for Technosignatures,** led by Shelley Wright, addresses the second goal of the workshop.

**Section 5, Emerging and Future Opportunities in Technosignature Detection,** led by Svetlana Berdyugina and Natalie Batalha, addresses the third and fourth goals of the workshop.

# 1.1 WHAT ARE TECHNOSIGNATURES?

The term “technosignatures,” a contraction of “technological signatures” or “signatures of technology,” first appears in the literature in an article by Jill Tarter (2007), who wrote,

> If we can find technosignatures—evidence of some technology that modifies its environment in ways that are detectable–then we will be permitted to infer the existence, at least at some time, of intelligent technologists. As with biosignatures, it is not possible to enumerate all the potential technosignatures of technology-as-we-don't-yet-know-it, but we can define systematic search strategies for equivalents of some 21st century terrestrial technologies.

This highlights several important aspects of the search for technosignatures that were stressed at the workshop and that should be kept in mind when reading this report:

1. Technosignatures are analogous to biosignatures in that they are a detectable sign of extant or extinct life. Note that while some consider technosignatures to be a subset of biosignatures and others think of them as being complementary to biosignatures, either way **searches for technosignatures are logically continuous with the search for biosignatures as part of astrobiology.**

2. **Technosignatures represent *any* sign of technology that we can use to infer the existence of intelligent life elsewhere in the universe**, including familiar objects of searches for extraterrestrial intelligence such as narrow-band radio signals or pulsed lasers. The term SETI (Search for Extraterrestrial Intelligence) often is used synonymously with the search for technosignatures.

3. As with biosignatures, one must proceed by hypothesizing a class of detectable technosignatures, motivated by life on Earth, and then designing a search for that technosignature considering both its detectability and its uniqueness. **The search for technosignatures is thus broad, encompassing much of astronomy.**

4. Unlike biosignatures, many proposed technosignatures are self-luminous or involve the manipulation of energy from bright natural sources. Also, since technological life might spread through the galaxy, its technosignatures might be found far in both space and time from its point of abiogenesis. **Compared to biosignatures, technosignatures might therefore be more ubiquitous, more obvious, more unambiguous, and detectable at much greater (even extragalactic) distances.**

# 1.2 WHAT ARE GOOD TECHNOSIGNATURES TO LOOK FOR?

Much discussion at the workshop and in the technosignature literature focuses on the merits of a particular proposed technosignature over others, often guided by the principle of looking for the overlap of what activities all civilizations are likely to participate in (such as using energy) and what signatures of those activities could be detected across interstellar distances. Many of these “axes of merit” were consolidated by Sofia Sheikh into a figure discussed on the final day of the workshop (see Figure 1).

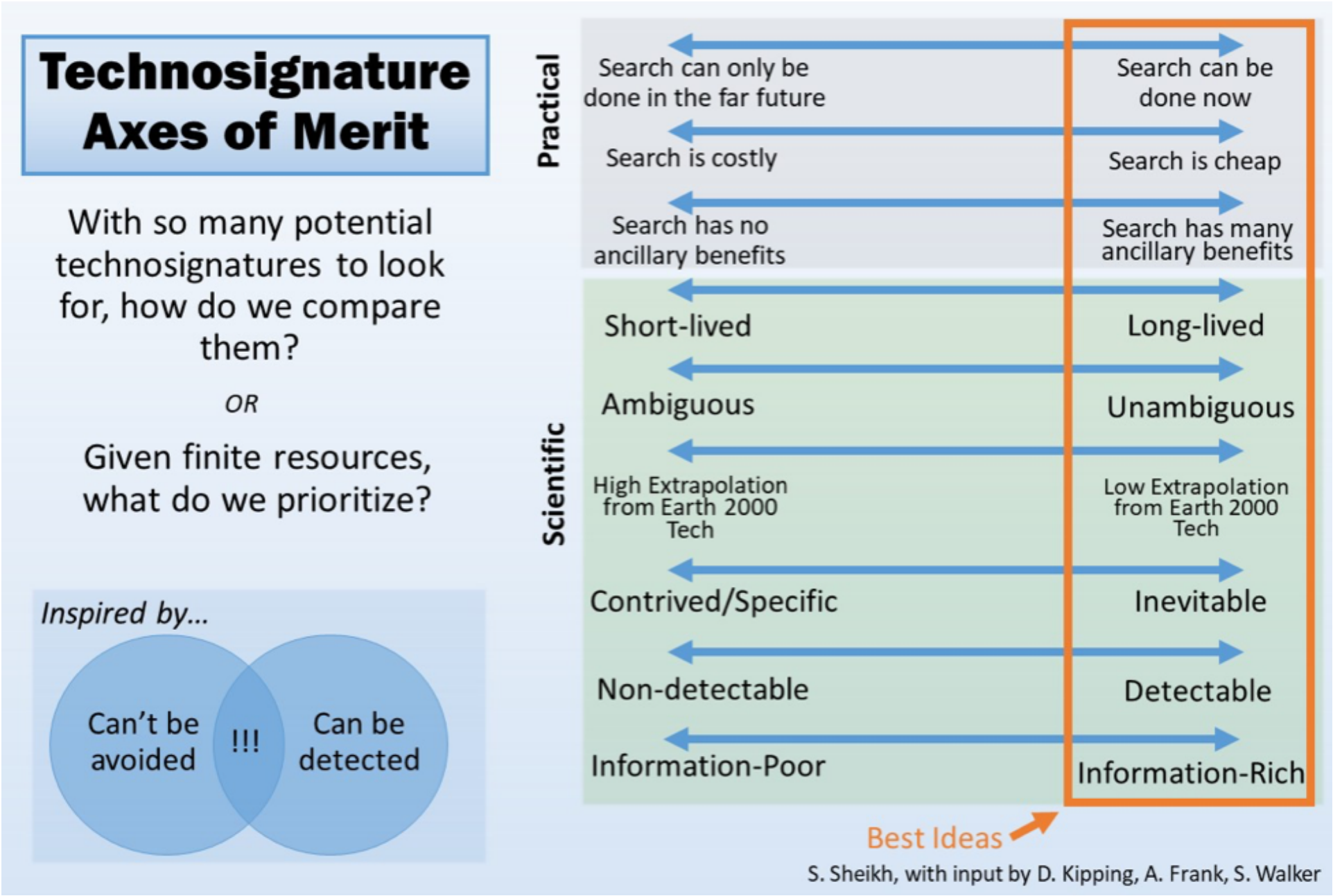

**Figure 1. Technosignature axes of merit, illustrating some of the considerations that go into developing a good search strategy for technosignatures.**

The axes illustrated in Figure 1 are:

1. The timeliness of the search, often driven by the amount of new technology needed to be developed or deployed in order to conduct a thorough search for a technosignature.

2. The cost of a search, which could include not only financial, but telescope time or other opportunity costs, given the availability of the technology.

3. The ancillary benefits of the search. Searches for many kinds of technosignatures involve surveys that can be used for other purposes, or can be expected to discover anomalies of significant and potentially transformative astrophysical importance. At the positive end of this axis are searches that satisfy Freeman Dyson’s “First Law of SETI Investigations”: “Every search for alien civilizations should be planned to give interesting results even when no aliens are discovered.”

4. The duration of the technosignature’s existence, or its duty cycle. For instance, signs of propulsion of interstellar craft might occur only in bursts, and so require long stares and repeated visits before being discovered; and might only occur for a brief period in a technological species’ development, and so only be present among a small fraction of host stars. Persistent technosignatures—such as ‘always on’ beacons, waste heat, or atmospheric pollutants—might be more ubiquitous and require only a single look.

5. The degree of ambiguity of a technosignature. As with biosignatures (such as $O_2$, which has both biotic and abiotic sources), some technosignatures might be easily mistaken for natural phenomena unrelated to life. For instance, waste heat from technology has a similar observational signature to astrophysical dust, which makes it difficult to distinguish, whereas certain narrow-band radio emissions may not have a natural source.

6. The degree of extrapolation from current Earth technology to generate the technosignature. We can be more confident that a technosignature might exist if we understand and deploy the underlying technology ourselves. For instance, laser and radio signals are popular targets of searches for technosignatures because humanity would be capable of detecting its own such signals at interstellar distances. Towards the other end of the axis, the creation of a Dyson sphere would be far outside humanity's current capabilities, and some proposed technosignatures (such as exotic forms of propulsion) require extrapolation beyond our best understanding of fundamental physics.

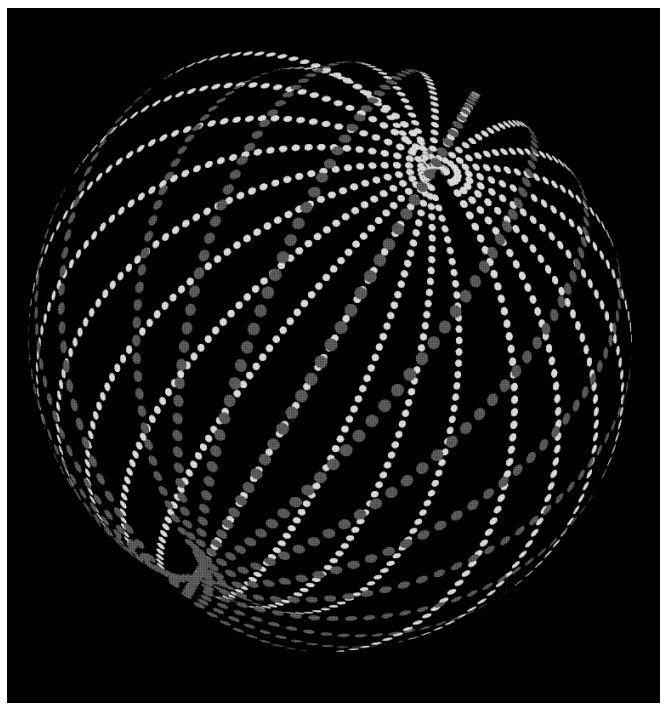

**A fanciful illustration of the concept of a partial Dyson sphere, or Dyson swarm. The concept of a hypothetical megastructure that encompasses a star and harnesses its energy, was first proposed by Freeman Dyson (1960). Image credit: Wikimedia Commons/ Vedexent.**

7. The degree of contrivance or inevitability of a technosignature. For instance, waste heat is an inevitable consequence of energy use according to fundamental physics, whereas the decision to send a transmission or to use a particular communication channel is a behavior whose probability cannot be predicted precisely, requiring extraterrestrial species to have a specific set of motivations and technologies.

8. The degree of detectability of the technosignature. As with biosignatures, a useful technosignature is a strong one. For instance, the spillover energy from a directed energy drive for interstellar spacecraft would be extremely bright and easily detected, whereas the transit signature of artificial satellites might be extremely subtle.

9. The utility of a positive detection. An information-rich technosignature, such as a deliberate message or the discovery of an extraterrestrial solar system artifact, would immediately enable intense follow-up and yield significant scientific gains. At the other end of the axis is a "bare contact" scenario, where we learn only that a technology exists at such great distance that follow-up would yield no additional gains.

There are interplays among these axes. For instance, many technosignatures become more detectable at the expense of being more contrived and requiring more extrapolation of current Earth technology, as with the increase in detectability of artificial structures in space with their size.

Many of the technosignatures discussed at the workshop were proffered because they optimized a figure of merit based on subjective weights given to these axes, especially the first two because the funding for such searches has been so low historically. In particular, at the extreme positive end of the first three axes are searches where the data collection of some fraction of a search volume has *already occurred* as part of surveys for other astronomical phenomena, and which only require robust upper limits to be calculated in the context of a technosignature search.

# 1.3 MATURITY OF THE FIELD

Because the field of searching for technosignatures has had no regular source of federal funding for decades, it is not as mature as other aspects of astrobiology, such as the hunt for biosignatures. In particular:

1. The literature on actual searches for technosignatures is rather thin compared to other aspects of astrobiology, and is dominated by a few mature search strategies that have enjoyed modest private funding.

2. Travel to conferences, publication of articles, and the advising of students is and will continue to be funding-constrained, since searches for technosignatures are not a topic covered by most existing federal grant money solicitations.

3. There is virtually no curriculum on the subject at the graduate level; fewer than 10 PhDs have been awarded for dissertations on searches for technosignatures.

4. As a result of items 1–3 above, there are few reviews of the field or metastudies to draw upon when addressing the goals of this workshop. The last review of the field was an *Annual Review of Astronomy and Astrophysics* (ARAA) article in 2001 by Tarter, which focused on radio SETI.

Thus, to a large extent, getting good answers to the questions posed by the workshop is not a matter of asking the appropriate experts to synthesize information that already exists, it will require training and supporting scientists to do the work necessary to generate that information in the first place.

A "meta" answer to the third and fourth questions ("*What new X would be important for future advances in the field? What role can NASA partnerships with the private sector and philanthropic organizations play in advancing our understanding of the technosignatures field?*) is that NASA has the opportunity to foster a scientific and educational community in which researchers from an interdisciplinary array of fields can be introduced to the history of research into technosignatures, the current state of progress of such searches, and learn the open issues in the field and where they can contribute to the future study of and search for technosignatures.

**For instance, regular meetings or workshops with strong didactic and review components and travel support, run by the few researchers with experience in searching for technosignatures, would generate a population of scientists with the background necessary to work on these questions.**

# 1.4 BREADTH OF THE FIELD

As with work on biosignatures, work on technosignatures is inherently interdisciplinary and will be unnecessarily hampered if it is siloed into a stand-alone program. As will be clear from the following sections, technosignature searches span the electromagnetic (EM) spectrum and beyond (into multi-messenger astronomy); span timescales from fast transients to persistent sources; cross disciplinary boundaries, including (to name just a few) those between traditional astronomy, information and network theory, planetary systems science, space- and ground-based instrumentation, astrobiology, risk communication, and public outreach; and span distance scales from the very nearby (the Moon and near-Earth environment) to the extragalactic (e.g., gravitational wave and fast transient sources at significant redshift).

Searches for technosignatures both inform and are informed by other areas of astronomy and astrobiology. Much astronomical data and instrumentation is dual-use, providing an opportunity for both searches for technosignatures and other areas of astronomy and astrobiology to contribute to the science goals and science output of missions. Candidate technosignatures are often inherently anomalous and interesting, requiring follow up and investigation in a traditional astronomical context.

**Searches for technosignatures are thus a potential motivating science case across NASA's science portfolio,** including most of its current, planned, and contemplated planetary science and astrophysics missions. They would also be natural topics for proposals to **most of its grant programs, its graduate and postdoctoral fellowship programs, and guest observer (GO) and data analysis programs for most of its observatories.** These synergies across disciplines and programs would provide NASA a rare opportunity for NASA to integrate a science goal into its science portfolio *from the beginning*.

## 1.5 LIMITATIONS OF THIS DOCUMENT

This report represents the contributions of those who were able to attend and afford the NASA Technosignatures Workshop on very short notice (less than 2 months in most cases) and who were also able to find time to volunteer to substantively contribute to and edit this report in a short timeframe (about 1 month). There are members of the technosignature search community who could not attend and many who did attend but did not have the time or resources to volunteer to the degree they would have liked.

## 1.6 AUTHORS OF THIS DOCUMENT

While the section leads named earlier in this report were primarily responsible for the content of their respective sections, all workshop participants were asked to contribute and many of their ideas presented or discussed at the workshop are reflected herein. They are:

| Participant Name | Organization | E-mail |
|---|---|---|
| Rana Adhikari | California Institute of Technology | rana@caltech.edu |
| Daniel Angerhausen | University of Bern | daniel.angerhausen@gmail.com |
| James Annis | Fermi National Accelerator Laboratory | annis@fnal.gov |
| Amedeo Balbi | University of Rome Tor Vergata, Italy | amedeo.balbi@gmail.com |
| Natalie Batalha | University of California, Santa Cruz | natalie.batalha@ucsc.edu |
| Charles Beichman | NASA Exoplanet Science Institute | chas@ipac.caltech.edu |
| Gregory Benford | Physics, University of California, Irvine | xbenford@gmail.com |

| Participant Name | Organization | E-mail |
|---|---|---|
| Jim Benford | Microwave Sciences | jimbenford@gmail.com |
| Svetlana Berdyugina | Leibniz-Institut für Sonnenphysik (KIS) / PLANETS Foundation | sveta@leibniz-kis.de |
| Anamaria Berea | Complex Adaptive Systems Lab, University of Central Florida | anamaria.berea@ucf.edu |
| Nathalie Cabrol | SETI Institute | Nathalie.A.Cabrol@nasa.gov |
| Steve Croft | University of California, Berkeley | scroft@astro.berkeley.edu |
| Paul Davies | Arizona State University | Paul.Davies@asu.edu |
| Ross Davis | Indiana University | rossdavi@iuk.edu |
| Gabriel De la Torre | University of Cadiz, Spain | gabriel.delatorre@uca.es |
| Julia DeMarines | Blue Marble Space Institute | julia.demarines@colorado.edu |
| Kathryn Denning | Anthropology Department, York University, Canada | kdenning@yorku.ca |
| Bill Diamond | SETI Institute | bdiamond@seti.org |
| Jessie Dotson | NASA Ames Research Center | jessie.dotson@nasa.gov |
| Jamie Drew | Breakthrough Listen | drew@breakthrough-initiatives.org |
| Emilio Enriquez | University of California, Berkeley | e.enriquez@berkeley.edu |
| Yan Fernandez | University of Central Florida | yan@physics.ucf.edu |
| Duncan Forgan | University of St. Andrews | dhf3@st-andrews.ac.uk |
| Adam Frank | University of Rochester | afrank@pas.rochester.edu |
| Dawn Gelino | NASA Exoplanet Science Institute | dawn@ipac.caltech.edu |
| Daniel Giles | Illinois Institute of Technology, Adler Planetarium | dgiles1@hawk.iit.edu |
| Eliot Gillum | SETI Institute | egillum@seti.org |
| David Grinspoon | Planetary Science Institute | david@funkyscience.net |
| Jamie Holder | University of Delaware | jholder@physics.udel.edu |
| Andrew Howard | California Institute of Technology | ahoward@caltech.edu |
| Albert Jackson | Lunar and Planetary Institute | al_jackson@aajiv.net |
| Shubham Kanodia | Pennsylvania State University | shbhuk@gmail.com |
| David Kipping | Columbia University | dkipping@astro.columbia.edu |
| Kevin Knuth | University at Albany (SUNY) | kknuth@albany.edu |
| Ravi Kopparapu | NASA Goddard Space Flight Center | ravikumar.kopparapu@nasa.gov |
| Eric J. Korpela | University of California, Berkeley | korpela@ssl.berkeley.edu |
| Jeff Kuhn | Institute for Astronomy, University of Hawaii and PLANETS Foundation | kuhn@ifa.hawaii.edu |
| Pauli Laine | University of JyvŠskylŠ, Finland | pauli.e.laine@jyu.fi |
| Glen Langston | National Science Foundation | glangsto@nsf.gov |
| Joseph Lazio | NASA Jet Propulsion Laboratory | Joseph.Lazio@jpl.caltech.edu |
| Philip Lubin | University of California, Santa Barbara | lubin@deepspace.ucsb.edu |
| Claudio Maccone | International Academy of Astronautics and INAF (Italy) | Claudio.Maccone@protonmail.com |
| Jean-Luc Margot | University of California, Los Angeles | jlm@astro.ucla.edu |
| Karen O'Neil | National Radio Astronomy Observatory | koneil@nrao.edu |

| Participant Name | Organization | E-mail |
|---|---|---|
| Paul Pinchuk | University of California, Los Angeles | ppinchuk@physics.ucla.edu |
| Chris Rose | Brown University School of Engineering | Christopher_Rose@brown.edu |
| Kevin Schillo | University of Alabama | kjs0011@uah.edu |
| Gavin Schmidt | NASA Goddard Institute for Space Studies | gavin.a.schmidt@nasa.gov |
| Megan Shabram | NASA Ames Research Center | mshabram@gmail.com |
| Sofia Z. Sheikh | Pennsylvania State University | szs714@psu.edu |
| Andrew Siemion | University of California, Berkeley | siemion@berkeley.edu |
| Steinn Sigurdsson | Penn State Extraterrestrial Intelligence Center (PSETI Center) | sxs540@psu.edu |
| Frank Soboczenski | King's College London | frank.soboczenski@gmail.com |
| Hector Socas-Navarro | Instituto de Astrofísica de Canarias | hsocas@iac.es |
| Andrew Stuart | Emory University | a.m.stewart@emory.edu |
| Jill Tarter | SETI Institute | tarter@seti.org |
| Doug Vakoch | METI International | dvakoch@meti.org |
| Sara Walker | Arizona State University | sara.i.walker@asu.edu |
| Dan Werthimer | University of California, Berkeley | danw@ssl.berkeley.edu |
| Pete Worden | Breakthrough Listen | pete@breakthroughprize.org |
| Jason Wright | Penn State University | astrowright@gmail.com |
| Shelley Wright | University of California, San Diego | saw@physics.ucsd.edu |

# 2 EXISTING UPPER LIMITS ON TECHNOSIGNATURES

Daniel Angerhausen, Emilio Enriquez, Andrew Howard, Shubham Kanodia, David Kipping, Ravi Kopparapu, Eric Korpela, Jean-Luc Margot, Andrew Siemion, Hector Socas-Navarro, Jason Wright

In 2005, Chyba and Hand wrote:

> Astro-physicists...spent decades studying and searching for black holes before accumulating today's compelling evidence that they exist. The same can be said for the search for room-temperature superconductors, proton decay, violations of special relativity, or for that matter the Higgs boson. Indeed, much of the most important and exciting research in astronomy and physics is concerned exactly with the study of objects or phenomena whose existence has not been demonstrated—and that may, in fact, turn out not to exist. In this sense astrobiology merely confronts what is a familiar, even commonplace situation in many of its sister sciences.

Their point holds just as well for searches for technosignatures. And like in those other fields, until a technosignature is discovered, progress will consist of:

1. Development of ways to hunt for technosignatures, and

2. Upper limits on certain kinds of technosignatures.

This section describes the second of these efforts. The first effort is discussed in the Section 3.

## 2.1 LIMITS AND THE LIMITATIONS OF LIMITS

The best upper limits are quantitative and rigorous, describing the kinds of signals that have been firmly ruled out by experiment or observation. Experiments that search for signals against little background can provide much cleaner upper limits than those that must measure or estimate their background, but in either case significant work must be done to translate a lack of detection (a “null result”) into a quantitative upper limit.

Consider, for instance, searches for radio technosignatures. An extremely narrowband astronomical radio signal can only have an artificial origin (because there are physical limits on the narrowness of natural coherent emission such as masers), so the discovery and confirmation of such a signal would be dispositive and immediately indicate the existence of extraterrestrial technology, and the lack of discovery in a given search means that those signals the search is sensitive to can be firmly ruled out. But such signals are also commonly generated by humans, and so all such searches are plagued by anthropogenic false positives in the form of radio frequency interference (RFI). Algorithms designed to excise RFI are imperfect and eliminate a fraction of the signals that could be indicative of technosignatures. What's more, we do not know what modulation scheme an extraterrestrial technology might employ, and so our sensitivity to such signals is a function of the algorithms used to analyze the radio telescope data.

Together, these problems make rigorous upper limits on radio technosignatures difficult. Comparison of results across different teams is also difficult: different teams make different assumptions about the signals they seek when designing their observational and data-reduction strategies, leading to different detection algorithms and thresholding methods, and different methods of RFI rejection and mitigation. Thus, it is challenging to compare null results across search programs.

There is also no agreed upon standard in the community on how to define an upper limit, both in terms of what statistical framework to employ and what confidence level to state. The community has not, for instance, shared data among teams to test their algorithms and reporting procedures, performed Monte Carlo injection-recovery tests to quantify their algorithmic sensitivities, or organized "data challenges" to compare overall methodologies.

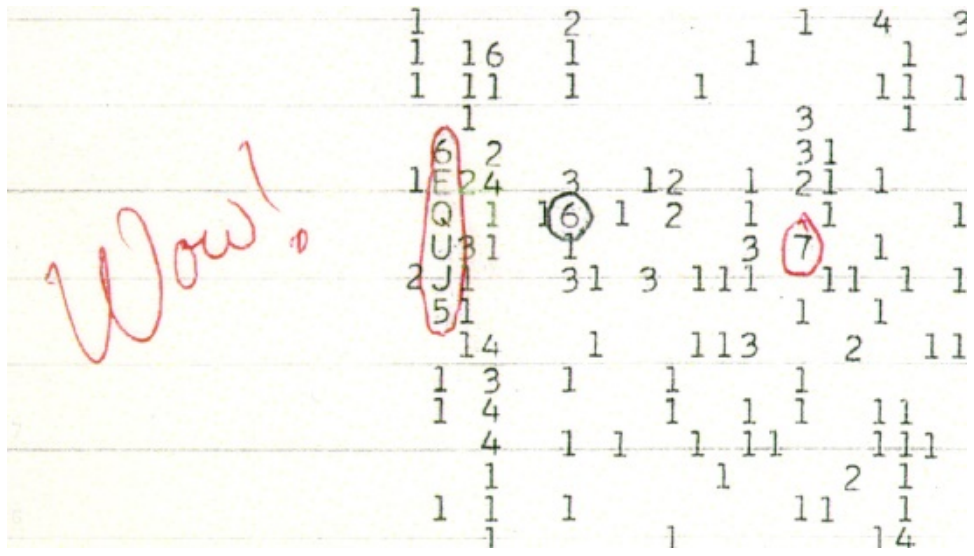

**On August 15, 1997, the Big Ear radio telescope at Ohio State University received a strong narrowband radio signal. Astronomer Jerry R. Ehman discovered the anomaly.**

Follow up also complicates upper limit calculations. Persistent or intermittent sources can be followed-up to determine their nature, but one-off detections (such as the notorious Wow! signal) are difficult to assess. For technosignatures with natural confounders (such as waste heat searches), upper limits can only be computed rigorously when the rate and nature of those confounders is well understood, which is not always the case.

There is also a major reporting problem. There is, of course, the usual 'file-drawer' problem of null results not being published because they are rarely cited and rarely accepted to prestigious journals. In addition, for the first several decades of technosignature searches, few journals would publish technosignature-related work at all. The historical lack of federal funding for the field also meant that much technosignature work also occurred outside of academia, and so outside of its publish-or-perish culture.

As a result, much of the literature on technosignature limits is contained in conference proceedings and unrefereed journals, if it appears anywhere at all, and so upper limits can only be deduced from approximations and guesses about a given survey's properties, algorithms, and the fact that they did not announce a detection.

Despite these difficulties, some quantitative upper limits have been computed, and others can be approximated, which are discussed in the following subsections.

# 2.2 UPPER LIMITS IN CONTINUOUS WAVE RADIO SEARCHES

Radio searches for technosignatures have been described in terms of searching for "needles" in the Cosmic Haystack—this emphasizes both the difficulty of the problem and how we seek signs of artifice among a huge set of natural confounders. The concept of the Cosmic Haystack was first developed quantitatively by Wolfe et al. (1981), and further by Tarter (2001) and Wright et al. (2018), among others.

The Cosmic Haystack has high dimensionality, reflecting the many ways we might search for a signal. Wright et al. (2018) chose the following nine dimensions, although other parameterizations are also used in the literature:

1. Power level of transmitter
2. Transmission central frequency

3-5. Distance and position (3 dimensions)

6. Transmission bandwidth
7. Time of transmission or repetition rate
8. Polarization of the transmission
9. Modulation scheme of the transmission

Upper limits can thus be expressed in terms of ruling out transmissions above a certain strength from a certain direction or star, at a given range of frequencies and transmission bandwidths, at certain times, and of a given modulation and polarization scheme.

Tarter et al. (2010) and Wright et al. (2018) both made attempts to estimate what fractions of the total space spanned by these nine dimensions have been searched for signals. They concluded that the total searched volume is small, calculating fractions of roughly $10^{-22}$ and $10^{-18}$, respectively. **These small numbers illustrate that the total searches completed to date is still quite small, and current upper limits are quite weak.** (It should also be noted that it is not necessary to search this entire space to make progress in the hunt for technosignatures; to use Jill Tarter's analogy: One does not need to search all of the world's oceans to find the first evidence of fish.)

Large gains remain to be made by sensitive, wide-field, broadband surveys in the future. There are many figures of merit that have been developed to compare the relative speeds of various searches of the Cosmic Haystack. Drake et al. (1984) described a commonly used figure of merit for narrowband radio searches:

$$\Delta\nu_{\rm inst}\Omega\phi_{\rm min,}^{-\frac{3}{2}}$$

i.e., proportional to the instrument's bandwidth, the sky coverage, and the minimum detectable flux (which has a -3/2 dependence because the minimum detectable signal drops as $d^{\frac{1}{2}}$, and the volume searched grows as $d^3$). Other figures of merit also exist, including those of Seeger and Wolfe (1985), Dreher and Cullers (1997), Harp et al. (2016), Enriquez et al. (2017), and Wright et al. (2018), each of which weighs haystack dimensions differently. Such figures of merit allow one to 'collapse' the Cosmic Haystack into a single dimension and roughly compare the relative search speeds (and so upper limits) of various surveys.

Enriquez et al. (2017) define the survey speed figure of merit, which is inversely proportional to

$$\frac{\Delta\nu_{\rm inst}}{\sigma^2\Delta\nu_{\rm chan}},$$

where $\Delta\nu_{\rm chan}$ is the bandwidth of a single channel, and $\sigma$ is the system equivalent flux density of the instrument (essentially its sensitivity in a single channel). Enriquez et al. then presented most of the previous large radio technosignature programs in a single plot using these two figures of merit (see Figure 2).

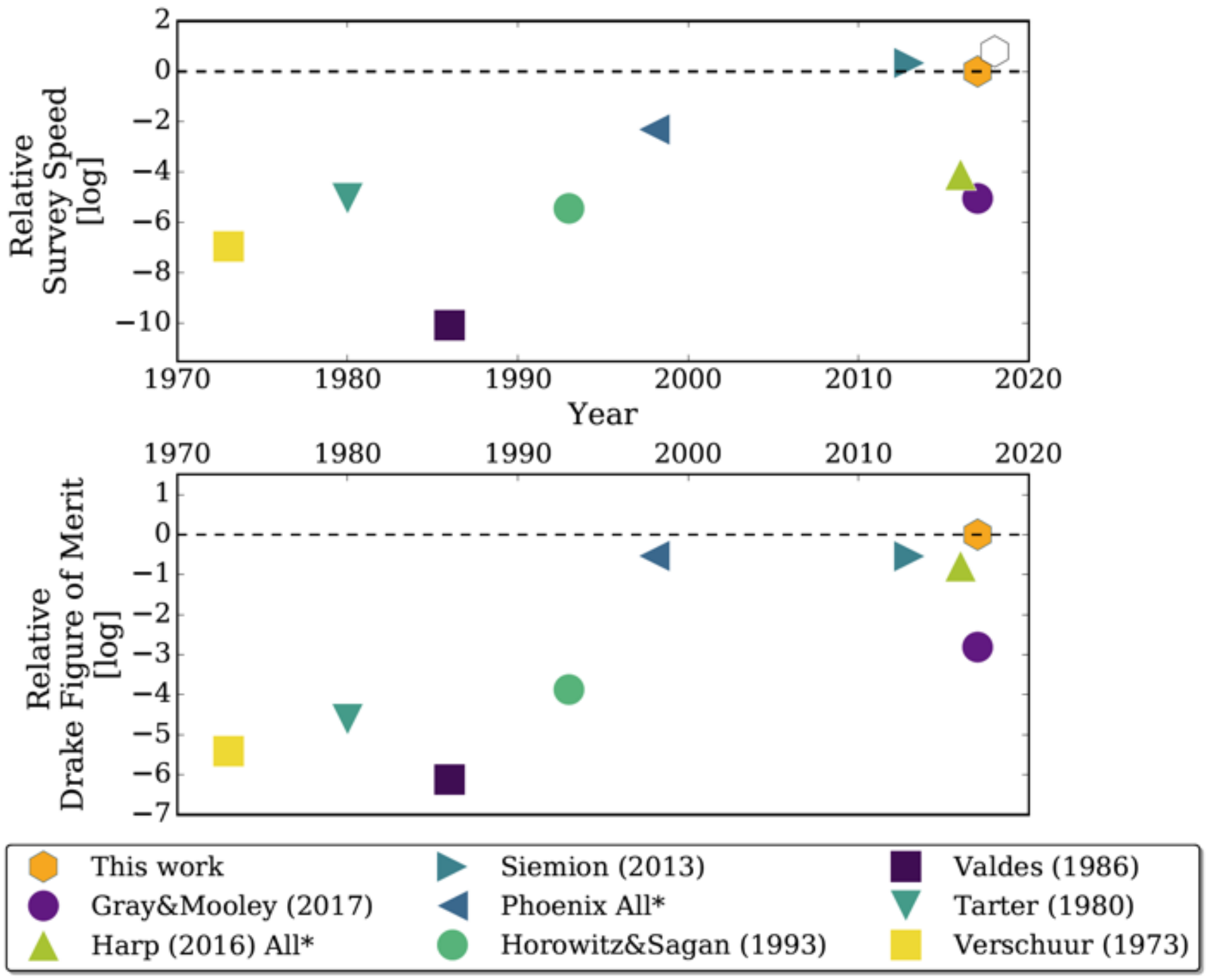

**Figure 2. Comparison of this work with several previous SETI campaigns. The top figure compares surveys based on relative survey speeds. The white hexagon takes into account the current instantaneous bandwidth available to the Breakthrough Listen backend ($\Delta\nu \approx 5$ GHz), which is underutilized in L-band observations. The bottom figure uses the relative DFM values for the comparison. Both figures only show the summed values for surveys with multiple components. Image and caption from Enriquez et al. (2017).**

In Figure 2, “this work” refers to the Breakthrough Listen Green Bank Observatory survey of 692 nearby stars in L-band. This gives a sense of which surveys have the largest haystack search speed by these figures of merit.

Enriquez et al. (2017) also described each search’s upper limits in terms of the transmitter rate, or the product of the number of stars searched and fractional bandwidth $\nu_{\rm rel}$, defined the instrumental bandwidth (i.e., the range of frequencies searched) divided by the central frequency searched (see Figure 3). They presented the quantity for various surveys as a function of the minimum detectable equivalent isotropic radiated power (EIRP; this is the total power of a transmission times the gain of the antenna; here, “gain” refers to the amplification factor above the isotropic value due to the beaming of the signal). This captures the tradeoffs between sensitivity and sky coverage in many surveys, and shows that the current limits set by the Breakthrough Listen project have a similar figure of merit to the surveys of the Andromeda galaxy by Gray and Mooley (2017), who were only sensitive to extremely powerful transmitters but surveyed hundreds of billions of stars.

Other key surveys described in these figures are Project Phoenix (Backus 1998, and others) as the largest scale survey before Breakthrough Listen, and surveys with the Allen Telescope Array (Harp et al. 2016), which extended the searches to higher frequencies. Recent surveys not included in the figures are: the low frequency surveys at the Murchison Widefield Array (Tingay et al. 2016, 2018), searches of interstellar asteroid `Oumuamua (Enriquez et al. 2017), and others (Margot et al. 2018).

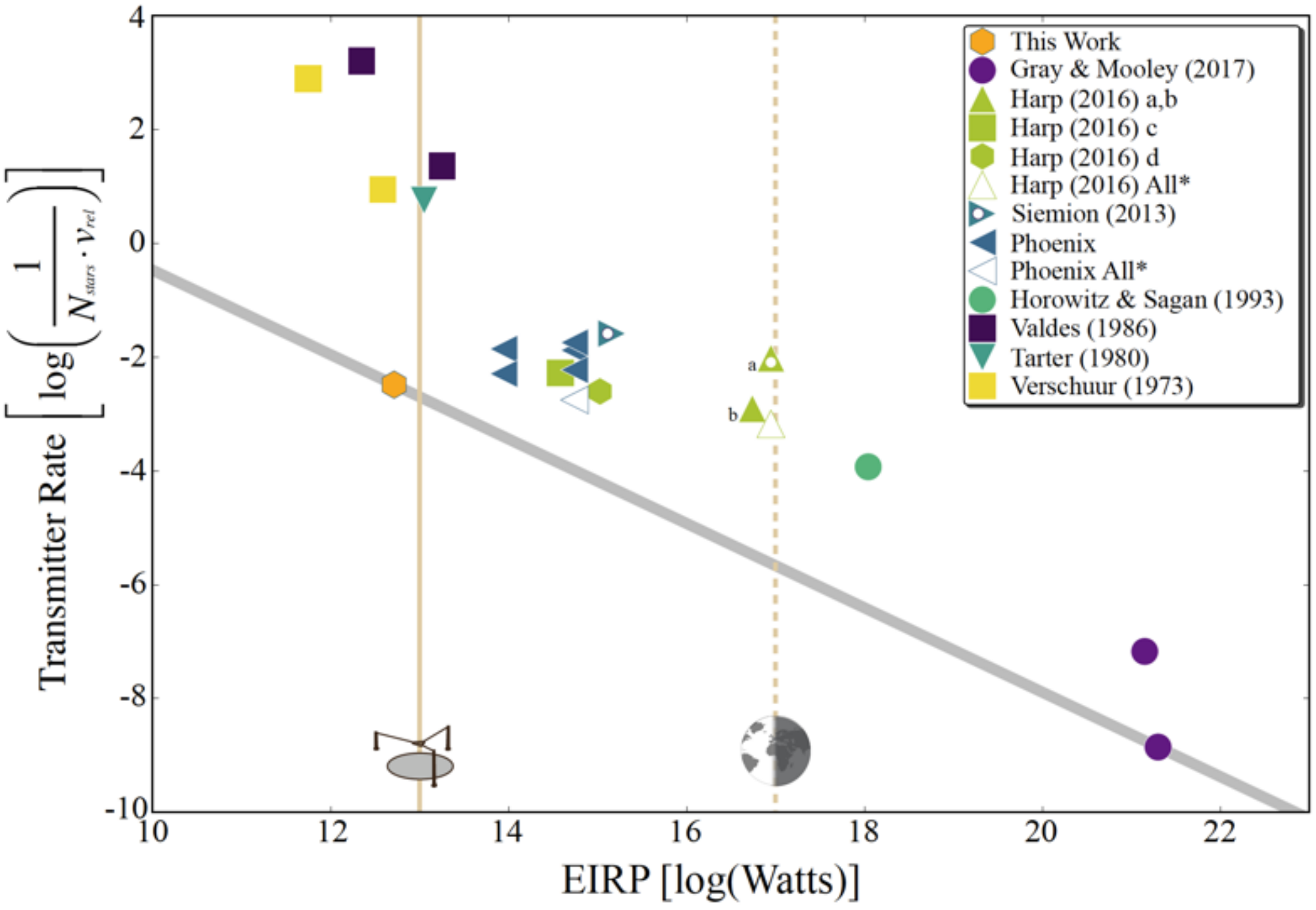

**Figure 3. Comparison of this work with several historic SETI projects. The vertical lines indicate characteristic EIRP powers, while the solid line indicates the EIRP of the AO planetary radar ($\mathcal{L}_{AO}$), and the dotted line indicates the total solar power incident on Earth's surface, commonly referred as the energy usage of a Kardashev Type I civilization ($\mathcal{L}$ ). The gray line is a fit of values for this work and that of Gray and Mooley (2017) by using Equation 12. The points labeled "All" show the total for a given project, this value is calculated by the sum of transmitter rates and taking the largest EIRP. EIRP values were calculated based on the most distant target for a given survey; sensitivity is better for nearer stars. The total for other works with multiple surveys are not shown for clarity since they lie right on top of the lowest point. The shapes used for the different surveys is related to the stellar spectral types. Shapes with more sides indicate surveys targeting a wider array of spectral types. Triangles are used for searches only looking at solar-type stars (FGK) and circles are used to denote sky surveys with more than just main-sequence stars. The triangles with a white dot in the center show surveys targeting known exoplanets in the habitable zone. Image and caption from Enriquez et al. (2017).**

## 2.3 UPPER LIMITS IN PULSED RADIO SEARCHES

While continuous wave radio searches identify the artificial nature of a signal by its frequency modulation or bandwidth, one can also compress signals in time, producing signals that are broadband but still easily distinguished from most known natural sources. Studies of the fast-transient radio sky searching for natural sources (e.g., gamma ray bursts, flare stars, pulsar giant pulses) thus also naturally search for such artificial signals their algorithms are sensitive to.

Lazio (2008) provides a good overview of current and emerging prospects in this domain, and von Korff et al. (2013) provides an overview of upper limits in this domain (also notable is the survey of Bannister et al. (2012) (see Table 1).

**Table 1. Survey parameters.**

| # | Author | Telescope | Year | Dedisp | Ref | $\upsilon_0$ (MHz) | σ | TO (K) | tsample (μs) | t(μs) |
|---|---|---|---|---|---|---|---|---|---|---|
| 1 | O'Sullivan et al. | Dwingeloo | 1978 | incoh | *a* | 5000 | 6 | 65 | 2 | 2700 |
| 2 | Phinney & Taylor | Arecibo | 1979 | incoh | *b* | 430 | 6 | 175 | $1.7 \times 10^4$ | $1.7 \times 10^4$ |
| 3 | Amy et al. | MOST | 1989 | incoh | *c* | 843 | 6 | - | 1 | $1.7 \times 10^4$ |
| 4 | Katz & Hewitt | STARE | 2003 | incoh | *d* | 611 | 5 | 150 | 125000 | 125000 |
| 5 | McLaughlin et al. | Parkes | 2006 | incoh | *e, f* | 1400 | 6 | 21 | 250 | 250 |
| 6 | Lorimer & Bailes | Parkes | 2007 | incoh | *g* | 1400 | 6 | 21 | 1000 | 1000 |
| 7 | Deneva et al. | Arecibo | 2008 | incoh | *h* | 1440 | 5 | 30 | 64 | 64 |
| 8 | Von Korff et al. | Arecibo | 2009 | coher | - | 1420 | 30 | 30 | 0.4 | 0.4 |

| # | Δν (MHz) | G (K $Jy^{-1}$) | Beam Ω | Beams | $t_{obs}$ (h) | $N_{pol}$ | sens (Jy μs) | $d_{max}$ (kpc) | Rate ($pc^{-3}yr^{-1}$) |
|---|---|---|---|---|---|---|---|---|---|
| 1 | 100 | 0.1 | $6.6 \times 10^{-6}$ | 1 | 46 | 1 | $2 \times 10^{-4}$ | 100 | $8.4 \times 10^{-8}$ |
| 2 | 16 | 27 | $6.6 \times 10^{-6}$ | 1 | 292 | 1 | 1300 | 250 | $8.7 \times 10^{-10}$ |
| 3 | 3 | - | $3.6 \times 10^{-8}$ | 32 | 4000 | 1 | $1.6 \times 10^{-5}$ | 22 | $5.4 \times 10^{-8}$ |
| 4 | 4 | $6.1 \times 10^{-5}$ | 1.4 | 1 | 13000 | 2 | $1.8 \times 10^{-9}$ | 0.21 | $1.6 \times 10^{-7}$ |
| 5 | 288 | 0.7 | $1.3 \times 10^{-5}$ | 13 | 1600 | 2 | 120 | 810 | $1.8 \times 10^{-13}$ |
| 6 | 288 | 0.7 | $1.3 \times 10^{-5}$ | 13 | 480 | 2 | 240 | 580 | $1.7 \times 10^{-12}$ |
| 7 | 100 | 10 | $8.1 \cdot\times 10^{-7}$ | 7 | 420 | 2 | 14 | 2300 | $9.0 \times 10^{-13}$ |
| 8 | 2.5 | 10 | $8.1 \times 10^{-7}$ | 7 | 1460 | 1 | 54 | 1200 | $1.6 \cdot\times 10^{-12}$ |

[a] O'Sullivan et al. (1978)
[b] Phinney & Taylor (1979)
[c] Amy et al. (1989)
[d] Katz et al. (2003)
[e] McLaughlin et al. (2006)
[f] Manchester et al. (2001)
[g] Lorimer & Bailes (2007)
[h] Deneva & Cordes (2008)
[i] MOST has 1 mJy of noise in each beam after 12 hours, http://www.physics.usyd.edu.au/sifa/Main/MOST

## 2.4 UPPER LIMITS IN OPTICAL/NEAR-INFRARED (NIR) LASER SEARCHES

Just as with radio emission, optical/NIR light can be compressed in the frequency and time domains to stand out as an obviously artificial and easily detected signal against astrophysical background (in this case starlight). Also, as with radio emission, such surveys often perform tradeoffs between sky coverage and sensitivity.

Most work in this domain has focused on pulses in the time domain, which require fast electronics and coincidence rejection. An alternative approach is to look for continuous laser emission by virtue of its compression in frequency space—that is, looking for laser emission lines in the high-resolution spectra of stars.

At the workshop, Andrew Howard prepared a summary slide describing the current upper limits in various domains.

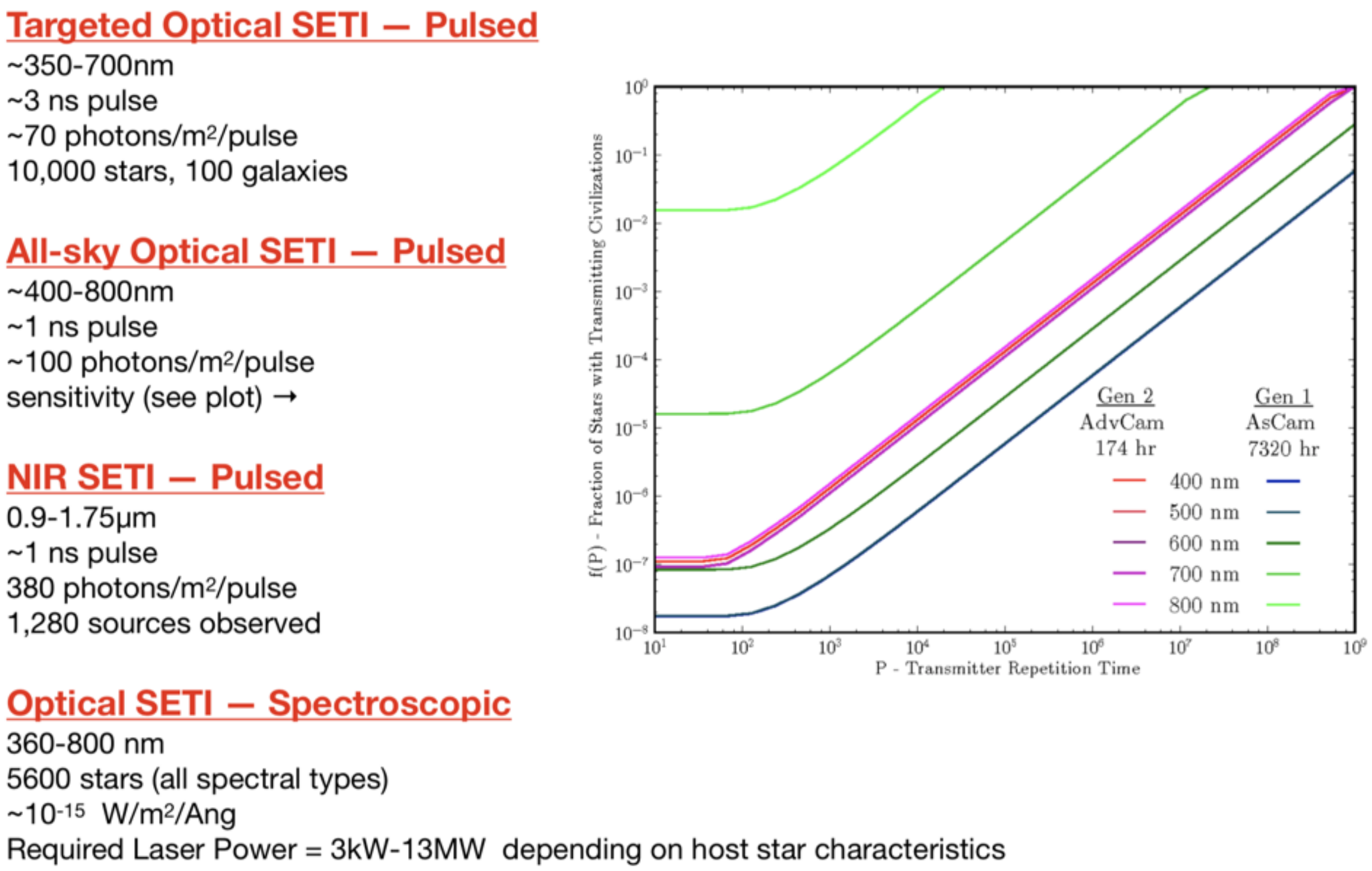

**Figure 4. Optical/NIR SETI—upper limits. Credit: Andrew Howard.**

More specifically:

- The Harvard survey of Howard et al. (2004) with sensitivity of 100 photons/m$^2$/pulse for a ~5 ns pulse in ~450–650 nm for 13,000 stars and a total observation time of 2,400 hrs. Based on survey details, upper limits were computed for the fraction, f(P), of stars with currently transmitting signals above the detection threshold that have a repetition period P. Example values of f(P) are f(1,000 sec) = $10^{-4}$ and f($10^7$ sec = 0.3 yr) = 1.
- Also from Howard et al. (2004), the Harvard+Princeton survey, which was a subset of the Harvard time, had simultaneous observations from a geographically separated instrument clone in Princeton. Using the same methodology as the Harvard targeted search, example values are f(1,000 sec) = $3\times10^{-2}$ and f($3\times10^5$ sec = 0.01 yr) = 1.
- Howard et al. (2007) and Mead (2013) describe a wide-field survey with sensitivity of ~70 photons/m$^2$/pulse in few ns pulses from 400–800 nm for the whole Northern Sky (declinations between -20 and 70 degrees), with an average dwell time of a few minutes. Using the same methodology as the Harvard targeted search, f(P) depended on wavelength, but was typically f(100 sec) = $10^{-7}$ and F(107 sec = 0.3 yr) = $10^{-2}$.

- The Berkeley and Lick surveys of Wright et al. (2001, 2014a SPIE 9147E0J), sensitive to 70 photons/$m^2$/pulse for a ~3 ns pulse in ~350–700 nm for 10,000 stars and 100 galaxies.
- The Very Energetic Radiation Imaging Telescope Array System (VERITAS) survey of Abeysekara et al. (2016), which achieves limits of 1 photon/$m^2$/pulse for pulses of durations between a few and a few tens of ns from 320–560 nm. They have published upper limits on 10 hours of integration on KIC 8462852, but have 10,000 hours on 20% of the sky in their archive.
- The Boquete (Panama) and Owl (Michigan) surveys of Schuetz (2018) and Schuetz et al. (2016), sensitive to 65 and 100 photons/$m^2$, respectively, for pulses of duration <1 ns to 50 ns from ~350–800 nm, with detectable pulse periodicities from 0.005 to >10 per second. Greater than 5,000 stellar observations with visual magnitudes of 3–14 and generally less than 300 ly distant. Both observatories are dedicated to optical SETI, and they are the foundation of Messaging Extraterrestrial Intelligence (METI) International's nascent optical SETI network.
- Tellis and Marcy (2015, 2017) report upper limits for 5,600 stars for laser lines from 364–789 nm with lasers of powers ranging from 3–3000 kW (dependent on the host star brightness and separation from the laser); and upper limits of 0.09–1 kW for sources angularly separated from 2,796 stars, including 1,368 Kepler Objects of Interest.

# 2.5 UPPER LIMITS ON WASTE HEAT AND STELLAR OBSCURATION

## 2.5.1 Dyson Spheres and Megastructures

Just as old as searches for radio emission from extraterrestrial technologies are searches for the waste heat of extraterrestrial industry. Dyson (1960) proposed that mid-infrared surveys could search for infrared excesses of nearby stars to detect the reprocessed starlight that might be captured by technology ("solar panels," effectively), and at least some of which, by fundamental laws of thermodynamics, must eventually be radiated as "waste" heat. Note that this waste heat does not imply anything about the *efficiency* of the underlying technology for doing its work, except in the case where that work is to generate coherent transmissions.

Upper limits in this domain are most conveniently expressed in terms of the AGENT formalism of Wright et al. (2014b), who parameterize the energy budget of a civilization in terms of its steady-state energy generation and radiation:

$$\alpha + \varepsilon = \gamma + \nu,$$

where

- $\alpha$ is the amount of starlight being intercepted
- $\epsilon$ is the amount of energy generated by other means

- $\gamma$ is the amount of energy being radiated at high entropy, as waste heat at some temperature $T_{\rm waste}$ set by the area of the radiating surface
- $\nu$ is the amount of energy being transmitted away at low entropy, such as coherent radio signals

and where all these quantities are normalized by the fraction of starlight available to a civilization. This parameterization is convenient because it allows one to quantify the upper limits one can place on the amount of waste heat or stellar obscuration from a source. Waste heat searches are then performed at infrared wavelengths, and obscuration is usually measured in the optical. In the extreme limit of $\alpha \to 1$, the swarm of structures blocking nearly all of the starlight of a single star is called a "Dyson sphere"; the very large structures orbiting a star intercepting its starlight are called "megastructures."

The fractional change in brightness of a star or galaxy caused by megastructures averaged over all directions is $-\alpha$ in the optical, but in the mid-infrared it is given by

$$\frac{\Delta F}{F} = \gamma \left( \frac{T_*}{T_{\rm waste}} \right)^3 - \alpha ,$$

where $T_*$ is the characteristic temperature of the starlight (where for most galaxies the red light is dominated by K giants). Thermodynamics requires that $\gamma$ be nonzero, and for starlight-fed industry that does not emit most of its energy supply at low entropy, we have $\gamma \approx \alpha$. Since for a large range of radiating surface areas, $T_{\rm waste} \approx$ 100−300 K, this shows that there is a large advantage to looking for waste heat in the infrared given by the large factor of $(T_*/T_{\rm waste})^3$ multiplying $\gamma$. This advantage is offset by the fact that many stars and galaxies have natural infrared excesses due to astrophysical dust, creating a large natural "background" against which signals of waste heat must compete.

### 2.5.2 Upper Limits on Megastructures in Transit

If megastructures exist orbiting stars, some of them should occasionally occult their host stars just as planets do. Kepler thus provided a natural search for megastructures in the course of its primary and secondary missions (K2), and these data need only be analyzed in a technosignature context to put upper limits on the sizes and orbits of megastructures around its target stars, much the way that exoplanets have been analyzed. Technosignatures in these data thus work against the twin backgrounds of spurious photometric variation (instrumental, shot noise, and intrinsic stellar variability) and natural occulters such as exoplanets as false positives.

Building on previous work by Arnold (2005), Korpela et al. (2015), Forgan (2013), and others, Wright et al. (2016) provided a framework for distinguishing technosignatures from these background sources. For ideal stars (brighter than 12th magnitude), Kepler had a sensitivity to approximately Earth-radius occulters on a timescale of hours. Of the ~200,000 Kepler stars observed, there are no clear examples of stars that cannot be explained naturally, although no formal upper limits have been calculated.

### 2.5.3 Upper Limits on Waste Heat from Stars

Weak limits on the existence of Dyson spheres or lesser degrees of stellar obscuration in the infrared have been set based on mid-infrared all-sky surveys. Jugaku and Nishimura (1991 and other papers in the series) used the Infrared Astronomical Satellite (IRAS) to perform a series of searches, and ultimately found no evidence of nearly-complete Dyson spheres around any of 545 sunlike stars within 25 pc of the Sun. Carrigan (2009) used a combination of IRAS photometry and spectroscopy to extend the search to the entire sky and found no good candidates among 11,000 sources with Low Resolution Spectrometer (LRS) spectra.

### 2.5.4 Upper Limits on Waste Heat from Galaxies

There have also been a few surveys of other galaxies that attempted to put upper limits on stellar obscuration. Annis (1999) studied 137 cluster galaxies looking for low-luminosity outlier of the Tully-Fisher and fundamental plane relations, and found no significant outliers (meaning none with $\alpha > 0.75$). Zackrisson et al. (2015) performed a similar study including infrared emission as a criterion and found no good candidates, and produced a tentative upper limit of 0.3% on the fraction of galaxies in the local universe with very large values of $\alpha$. Griffith et al. (2015) used the Wide-field Infrared Survey Explorer (WISE) and Two Micron All-Sky Survey (2MASS) to put upper limits on $\gamma$ among 100,000 galaxies resolved by WISE. They found that none had $\gamma > 0.85$ and only 50 were consistent with $\gamma > 0.5$, all of those presumably being starburst galaxies whose infrared emission came from dust in star-forming regions.

## 2.6 UPPER LIMITS ON SOLAR SYSTEM TECHNOSIGNATURES

Technosignatures in the solar system might come in the form of free-floating probes or structures—either passing through the solar system or in orbit around the Sun or other body—or in the form of structures or other signs of technologies on planetary surfaces.

The advent interplanetary probes placed the first weak upper limits on the existence of extraterrestrial technology on the Moon, Mars, and other solar system bodies: despite the claims of Lowell and others, these surfaces obviously have no evidence of cities or other large, kilometer-scale technology. Despite the time since these discoveries, the quantification of even this obvious upper limit has not yet been made. Newer high-resolution imagery of the Moon and Mars provide an opportunity for orders-of-magnitude improvement on these upper limits—one such program is that of Davies and Wagner (2013).

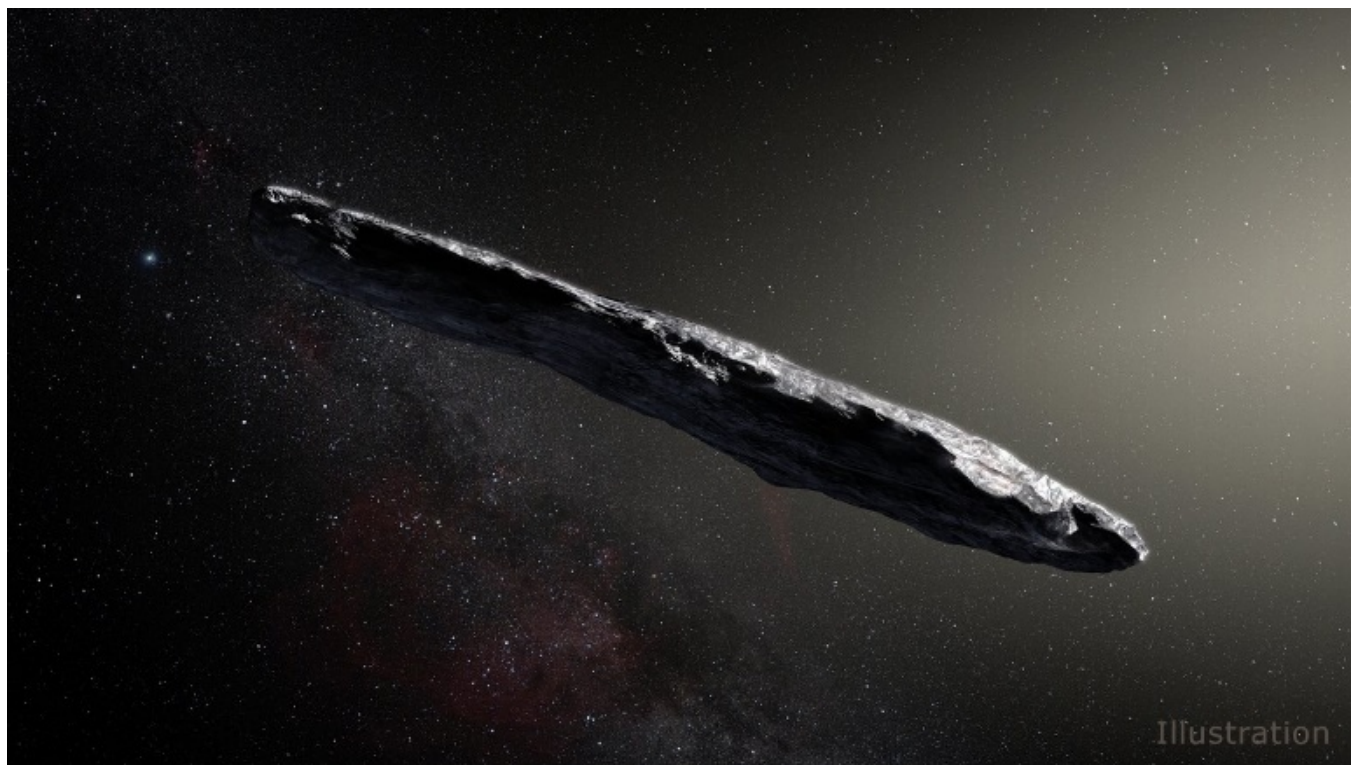

**Artist's concept of interstellar asteroid 1I/2017 U1 (`Oumuamua) as it passed through the solar system after its discovery in October 2017. Image credit: European Southern Observatory / M. Kornmesser**

There is also work to compute upper limits for chemical and other technosignatures on the ancient Earth. Industrial signatures might encompass radioisotope anomalies, presence of synthetic chemicals (chlorofluorocarbon (CFCs)), plastics, long chain persistent organic pollutants (POPs;

etc.), and combinations of unusual (though non-unique) geochemical markers (O,C,N stable isotopes, spikes in metal concentrations etc.); these have been computed by Schmidt and Frank (2018).

Work on free-floating technology is limited. For interstellar objects in transit through the solar system, the best upper limits are set by searches for interstellar asteroids, of which exactly one is known (`Oumuamua)—the upper limits on interstellar artificial objects is thus somewhere below the limits on those, but have not been quantified.

For objects in orbit around the Sun or other solar system objects, the only searches to date are those by Freitas and Valdes (2015 and earlier papers), who ruled out objects with geometric albedo > 0.1 and sizes >1–10 m in the Earth-Moon Lagrange points.

Although neither form of technosignature—free-floating or on surfaces—has strong quantitative upper limits yet, Haqq-Misra and Kopparapu (2012) have provided a Bayesian framework for such a calculation.

# 3 State of the Art of Technosignature Searches

Daniel Angerhausen, Emilio Enriquez, Andrew Howard, Shubham Kanodia, David Kipping, Ravi Kopparapu, Eric Korpela, Jean-Luc Margot, Andrew Siemion, Hector Socas-Navarro, Jason Wright

## 3.1 RADIO TECHNOSIGNATURES

Radio[1] technosignatures could take many forms, from the deliberately engineered narrowband transmissions sought in the first SETI projects, to pulses characteristic of (hypothetical) direct energy propulsion systems for interstellar spacecraft. As with work at other wavelengths, detection of such technosignatures requires several components:

1. Antennae and dishes for the collection of radio signals
2. Analog and digital backends that filter and record frequencies of interest
3. Algorithms, computational techniques, and hardware for storing, distributing, and analyzing the data for technosignatures

The first of these can take many forms but is usually generic in the sense that radio telescopes designed for technosignature searches will also serve well as instruments for at least some aspects of radio astronomy generally. The second requires more specialization—detection of extremely narrowband or fast signals requires hardware similar to that used in radar astronomy, pulsar astronomy or the detection of fast radio bursts, but rarely needed for other aspects of radio astronomy. Data sets from these instruments are often extremely large, as they are capable of generating data at least as quickly as modern computation can process it. These data rates put great demands on the third of these components, and indeed the state of the art in distributed computing generally is based in large part on methods used to analyze data in the search for radio technosignatures: the groundbreaking and widely used Berkeley Open Infrastructure for Network Computing (BOINC) distributed computing platform was developed for the SETI@Home project.

Because funding and telescope time allocations for radio technosignature work has historically been so low, a common scheme for radio technosignature work is to “share” time on a telescope, conducting a technosignature search at the same time as other science is being done. This is called “commensal” observing. There are several ways to accomplish this:

- A second instrument can sit at a different part of a dish’s focal plane, observing a different part of the sky from the primary instrument and doing its own science. In such cases, the second instrument will necessarily be looking at a different part of the sky than the primary instrument, so its pointings are to some degree “random.”

[1] The low-energy region of the electromagnetic spectrum has many components, including the submillimeter, microwaves, and longer wavelengths. Here we use the generic term “radio” to refer to all of these.

- Unlike with instruments at higher frequencies, which destroy photons to detect them, most radio telescopes measure voltages, and so can “share” their data among many instruments without loss of signal strength. This allows many backends to simultaneously filter and record signals from the same receiver. For single-dish, single-receiver instruments this allows the second backend to observe the same directions as the primary instrument.
- With interferometric arrays, the beam-forming allows backends to select which portions of the primary beam (i.e., the field of view of a single dish, typically quite large) are analyzed. Rather than imaging the entire field, many projects save computational time by analyzing only a few directions of interest (“beam forming”), such as nearby stars. These computational savings can then be applied to improving the bandwidth or spectral resolution of the measurements. Commensal work at telescope arrays is thus able to have more control over which targets are observed by forming beams at targets of interest within the primary beam, even if these are not the targets of interest of the primary instrument.

### 3.1.1 Existing Telescopes and Telescope Arrays

Although in principle, the entire suite of radio telescopes around the world could be put to use for radio technosignature work, there are only a few that have done such work recently. They are:

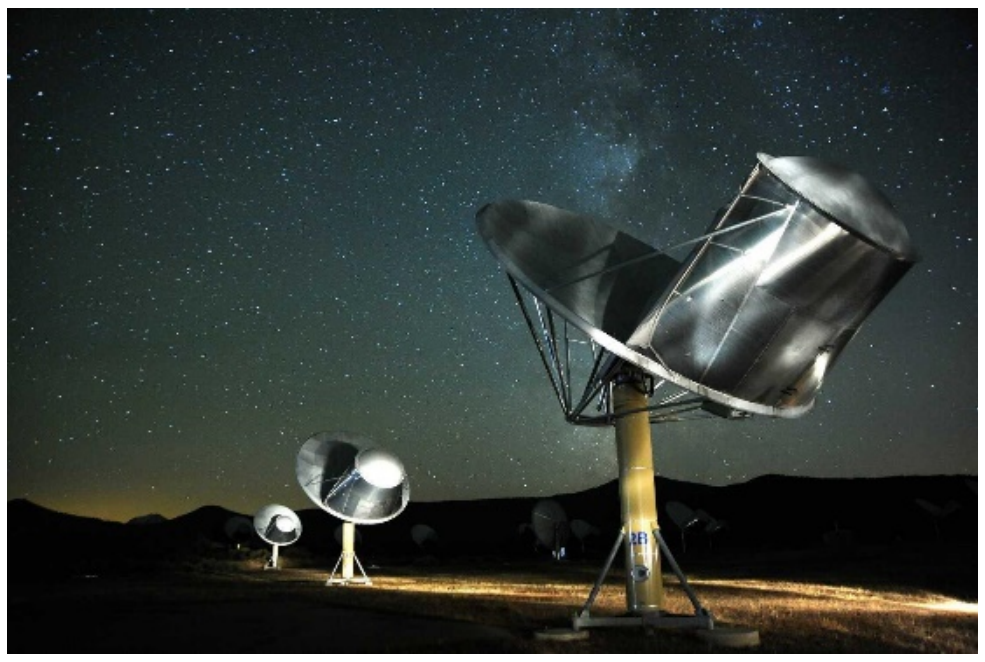

**Allen Telescope Array, built and operated by the SETI Institute. Credit: Seth Shostak/SETI Institute.**

- The Allen Telescope Array, built and operated by the SETI Institute primarily to conduct searches for radio technosignatures. This is an interferometric array with 42 small (6 m) dishes, designed to work across a wide range of frequencies and with a wide field of view.
- The Robert C. Byrd Green Bank Telescope, a 100 m single-dish telescope. This telescope currently has a significant fraction of its time used by the Breakthrough Listen initiative to conduct sensitive targeted searches at very high bandwidth (searching many GHz simultaneously at a few Hz spectral resolution). This is accomplished through a custom backend built on site. Radio technosignature work is also performed here by the University of California, Los Angeles (UCLA) SETI Group using the Green Bank Ultimate Pulsar Processing Instrument (GUPPI) backend designed for pulsar work, and historically commensal work has been done using Search for Extraterrestrial Radio Emissions from Nearby Developed Intelligent Populations (SERENDIP) backends.
- The 64 m Parkes Telescope in Australia, a single-dish, multi-receiver telescope also used by the Breakthrough Listen Initiative to search the Southern Sky for radio technosignatures. This is accomplished through a custom backend built on site.
- The 300 m Arecibo Telescope can point within 20 degrees of zenith. Radio technosignature work here is done by the UCLA SETI Group during dedicated time using the PUPPI backend, and commensally with the long-running SERENDIP projects. Some data is analyzed in both the frequency domain (via SETI@Home or other software) and the time domain (via Astropulse or other software).

- The Very Large Array can be and has been used for occasional dedicated radio technosignature work, including studies of nearby galaxies (Gray and Mooley 2017).
- The Murchison Widefield Array is a low-frequency (80–300 MHz) array of 2,048 antennae in western Australia. It is a low-frequency precursor to the Square Kilometer Array with a wide field of view and large collecting area (512 $m^2$). Some of its data have been studied in a radio technosignature context (Tingay et al. 2016, 2018).
- Low-Frequency Array (LOFAR) is a network of low frequency telescopes in Europe with a very large collecting area ($3\times10^5$ $m^2$) with the capability of performing very wide-field searches for radio technosignatures. Telescope time has been allocated in the past for SETI work (Enriquez et al., in prep.).
- Many European radio telescopes are involved in occasional SETI work, in particular the 64 m Sardinia Radio Telescope.

### 3.1.2 Existing Analysis Methods and Hardware

Radio surveys come in two primary types, the targeted searches with target specific stars, galaxies, or other celestial objects; and the sky survey searches, which cover large areas of sky rather than choosing specific targets. In radio/microwave SETI, the discrete Fourier transform is the dominant technique for converting the voltages recorded at the telescope into a spectrum that can be searched for either narrowband or broadband signals.

Because the sensitivity of a spectrometer to narrowband signals is proportional to the instrument resolution, narrowband SETI experiments divide the spectrum into a large number of narrow channels, about 1 Hz in width. At widths less than this, the drift of the signal due to Doppler shift due to the motion of the Earth becomes comparable to the channel width, which reduces the sensitivity. The first such commensal project, SERENDIP, was deployed in 1979 and had 100 channels and covered a 50 kHz bandwidth. The current generation, SERENDIP VI, has more than 4 billion channels and covers 3 to 5 GHz of bandwidth (see Table 2). These systems are custom field programmable gate array (FPGA)-based spectrometers.

**Table 2. SERENDIP programs.**

| Program | Year | Channels | Channel Width (Hz) | Instantaneous Bandwidth (MHz) | Full Bandwidth (MHz) | Band (MHz) | Sensitivity ($W/m^2$) | Sky Coverage | Relative Drake-Gulkis FOM |
|---|---|---|---|---|---|---|---|---|---|
| SERENDIP | 1979 | 100 | 500 | 0.05 | 20 | 1612 | $5\times10^{-22}$ | Small | ~0 |
| SERENDIP II | 1986 | 64 ki | 0.98 | 0.0647 | 2.5 | Various | $4\times10^{-24}$ | ~0.02 | 1 |
| SERENDIP III | 1992 | 4 Mi | 0.60 | 2.5 | 12 | 430 | $3\times10^{-25}$ | 33% | 3900 |
| SERENDIP IV | 1998 | 168 M | 0.60 | 100 | 100 | 1420 | $6\times10^{-25}$ | 33% | 11400 |
| SERENDIP V.v | 2009 | 128 Mi | 1.5 | 200 | 200x14 | 1420 | $4\times10^{-25}$ | 33% | 42000 |
| SERENDIP VI | 2014 | 4260M | 0.8 | 3408 | 243×14 | 1420 + 357 | $3\times10^{-25}$ | 33% | 78000 |
| SETI@home | 1999 | 62M×75k | 0.074-1220 | 35 | 2.5×14 | 1420 | $3\times10^{-26}$ | 33% | 25000 |

A consortium has been developed for developing these and other related instruments for radio astronomy, CASPER, the Collaboration for Radio Astronomy Signal Processing and Electronics Research. This collaboration includes corporate, university, and government partners and is funded by both government and private/corporate funding.

The Breakthrough Listen systems have comparable capabilities than the SERENDIP systems (GHz bandwidths with billions of channels), although built from commodity hardware including rackmount computers and consumer-grade graphics processing units (GPUs). These systems allow for greater flexibility on the types of data stored that could be used for other sciences.

A technique known as coherent dechirp can remove the 1 Hz channel width limit, but this method is computing intensive. The SETI@home project uses this method to achieve nearly a factor of 10 increase in sensitivity over the standard spectrometers described above ($3{\times}10^{-26}$ W/m$^2$) SETI@home is a “volunteer computing” project. Volunteers download the SETI@home software and analyze data that was recorded at radio telescopes. Currently SETI@home records data at the Arecibo Observatory, and also receives data from the Breakthrough Listen recorders at the Green Bank Telescope (GBT) and the Parkes Telescope in Australia.

For SETI@home, transmission bandwidth and storage space at the observatories limit processing to about 1% of the total Breakthrough Listen data. In addition to searching through 59,000 possible Doppler drift rates for narrowband signals, SETI@home performs beam profile fitting to help reject terrestrial signals, searched for pulsed signals on timescales from 1 ms to 1 minute, and uses an autocorrelation method to look for arbitrary repeated waveforms that might indicate technology.

Currently 120,000 people worldwide run SETI@home, providing 360 trillion floating point operations every second, equivalent to a large supercomputer. Since SETI@home started in 1999, over 10 million people have downloaded and run the SETI@home software. The success of SETI@home led to the development of the BOINC, which enables researchers to create other volunteer computer projects. BOINC development was funded through the National Science Foundation (NSF). Currently, there are more than 50 active projects using this infrastructure for cancer research, protein folding, atmospheric simulations, searching for gravitational waves, and other compute-intensive science projects.

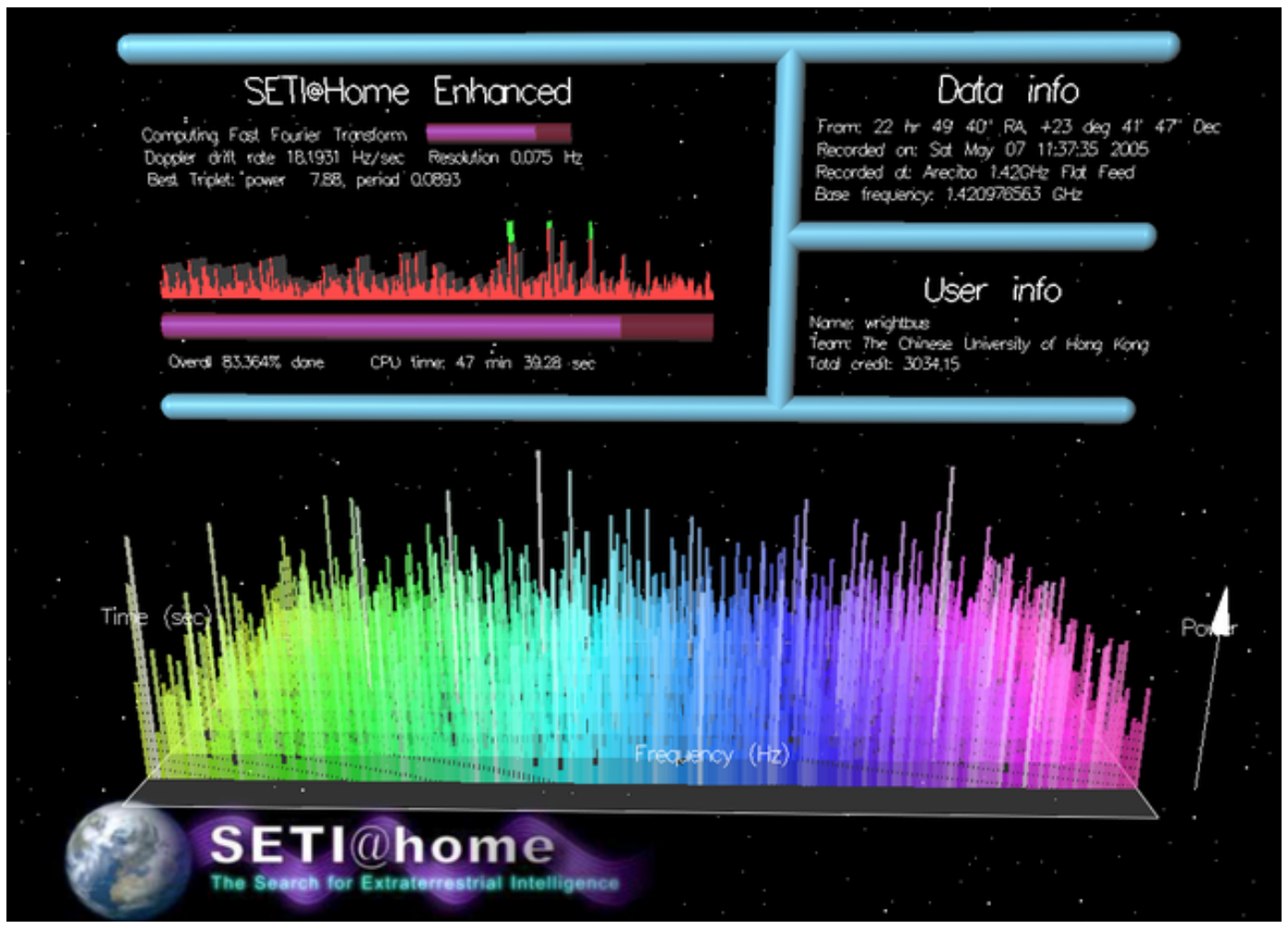

**SETI@home Enhanced BOINC screensaver. Credit: LGPL, https://commons.wikimedia.org/w/index.php?curid=1013154**

Because signals will drift due to motion of the Earth and possible motion of the transmitter, a search through possible Doppler drift rates is performed. In most experiments, this drift is performed using a tree search. Current searches for narrowband signals in Breakthrough Listen use TurboSETI,[2] an algorithm that performs incoherent corrections for Doppler accelerations. In the SETI@home experiment, which has very high compute capability, the Doppler search is performed using a coherent Doppler correction. This method allows SETI@home to use a 0.074 Hz channel width, which increase sensitivity by a factor of about 10.

Broadband SETI searches do not suffer from the same problem, as the Doppler drift is less than the width of the signal. However, broadband SETI searches do suffer from interstellar dispersion, which slows low frequencies relative to high frequencies. A tree search is typically used to search for dispersed signals. A coherent dedispersion algorithm can be used, but it demands far more computational resources. Thus far it has only been deployed in SETI in the Astropulse survey.

# 3.2 OPTICAL/NIR LASER TECHNOSIGNATURES

Work on laser technosignatures is almost as old as work on radio and waste heat technosignatures, tracing its origins to Schwartz and Townes (1961). Although there is less work here than in radio, it is nonetheless one of the few mature search strategies. Laser technosignatures might be compressed in time or frequency space, and the two approaches require different instrumentation. Continuous wave lasers can be detected with conventional high-resolution spectrographs, and so can be performed in principle on any new or archival spectral data (especially those of nearby stars thought to be candidates for life-bearing planets). Pulsed laser searches typically require very fast, custom photon-counting hardware with good coincidence detection to reject spurious signals.

## 3.2.1 Existing Projects

There are multiple ongoing efforts in the pulsed domain including:

- The Near-Infrared Optical SETI (NIROSETI) at the 1 m Nickel telescope at University of California Observatories (UCO)/Lick Observatory is searching nearby stars for pulsed infrared signals.
- VERITAS is a Cherenkov telescope designed for cosmic ray detection whose data can also be interpreted in a pulsed optical technosignature context. VERITAS has very large aperture (nearly 500 $m^2$) and field of view (3.5°) at the expense of angular resolution. Thirty (30) hours of dedicated observations of Breakthrough Listen targets are planned for 2018/2019. Archival data covers a significant fraction of the Northern Sky.
- The Harvard All-Sky Optical SETI program is a very wide-field pulsed optical SETI survey with a custom telescope and backend.
- The Boquete Optical SETI Observatory (0.5 m Newtonian) has been in continuous operation by METI International since 2010. The collaborative effort of Boquete and Owl Observatories

[2] https://github.com/UCBerkeleySETI/turbo_seti

in photometer development has made significant advancements in expanding the detection parameters.

- The Owl Observatory (0.4 m Cassegrain) is equipped with a photometer similar to that of Boquete. It has been in periodic, short-term operation for years, but will soon be operating for extended periods by METI International.
- The UC Santa Barbara Trillion Planet Survey uses the Las Cumbres Observatory telescope network and other efforts to perform differential imaging of nearby galaxies to search for flashes consistent with directed energy propulsion systems. (Lubin 2016, Lubin et al. 2016, Stewart et al. 2017).

There are two ongoing continuous wave laser searches:

- The Automated Planet Finder telescope at UCO/Lick Observatory measures high-resolution spectra of potentially planet-bearing stars; a custom data reduction pipeline operated by the Breakthrough Listen Initiative searches these spectra for unresolved emission lines consistent with a point source on the sky.
- Similarly, Keck/ High Resolution Echelle Spectrometer (HIRES) measures high-resolution spectra of potentially planet-bearing stars and typically employs a tall slit that explores the sky for a few arcseconds around the stars as well; a custom data reduction pipeline operated by the Breakthrough Listen Initiative searches these spectra for unresolved emission lines consistent with a point source on the sky.

## 3.3 ALGORITHMS AND SEARCH STRATEGIES

For optical technosignatures, the small stellar sampling and various search method inadequacies of the past 20 years are by far the better reasons for null results than the possibility that we are not being targeted with laser-like signals. The techniques used in all searches yield results only for the methods employed and for the specific time interval. Detection projects of the future need to have the broadest set of search parameters and the longevity to search the stellar neighborhood out several hundred light years.

A study of the future potentials for space and Earth based high energy lasers is needed to better define our search parameter boundaries. The study should include optical band, pulse repetition rate, targeting rate, received pulse periodicity, and laser longevity. Note that the received pulse periodicity is a consequence of all of those factors plus other technical and economic aspects. The study may also provide a basis for the calculation of a useful figure of merit for optical technosignature searches.

## 3.4 SEARCHES FOR ASTROPHYSICAL EXOTICA

Searches for technosignatures typically proceed by hypothesizing a particular technosignature and conducting a search for it against a background of natural or instrumental confounders. Another approach is to assume that strong technosignatures are very rare (but exist) and seek arbitrary rare phenomena unanticipated by theory for any natural source. At the Houston workshop, David Kipping demonstrated an algorithm ("the weird filter") for identifying certain classes of anomalous transits and another for inverting light curves to deduce the shape of occulters, and Walkowicz et al. (2014) have discussed using machine learning to mine light curves for technosignatures in an "agnostic" way.

One ongoing project for such exotica is Vanishing & Appearing Sources during a Century of Observations (VASCO) (Villarroel et al. 2016), a search for "disappearing stars" or other exotica in the time domain in existing archives of all-sky photometric surveys. Such work is essential preparation for the significantly more powerful data sets that Large Synoptic Survey Telescope (LSST) will provide in the near future.

## 3.5 OTHER ONGOING SEARCHES

Because funding for searches for technosignatures is so scarce, many searches are done on a "spare time" basis, and so do not have a formal structure or timeline for completion. A good example is the work of Carrigan (2009) who analyzed IRAS photometric and spectroscopic data for evidence of Dyson spheres. IRAS launched in 1983, the results of the photometric component of the search was announced in 2004, and the spectroscopic results were published in 2009, both as single-author works. There are many similar projects moving at a slow pace across the technosignature search landscape, so this section should not be considered comprehensive, or even representative, of the state of the art in all fields.

# 4 CURRENT AND NEAR-TERM ASSETS AND PROJECTS IN THE TECHNOSIGNATURE FIELD

Rana Adhikari, Charles Beichman, Anamaria Berea, Yan Fernandez, Daniel Giles, Joseph Lazio, Karen O'Neil, Matthew Pasek, Megan Shabram, Douglas Vakoch, Shelley Wright

## OVERVIEW

In principle, the technosignature field spans all wavelengths across the electromagnetic spectrum and includes non-EM communication that may include neutrinos, unique particles, and potential gravitational wave signatures. Current technosignature experimental missions primarily focus on radio wavelengths with a few dedicated unique technosignature instruments at optical wavelengths. Given limited resources, these technosignature studies have been performed with only a handful of scientists globally. Yet the astronomical community possess a wealth of current facilities, which if used for unique observations or for shared data/commensal operation, would greatly expand the technosignature field. This section provides a brief outline of existing facility-class multiwavelength observatories and instruments that could be used for the technosignature field, as well as existing or near-term dedicated technosignature instrumentation and facilities.

There are synergies between the astrobiology and exoplanet missions for biosignatures that have a particular relevance to the technosignature field, which are not currently encapsulated in today's research portfolios. Additionally, this section highlights facilities and surveys that have auxiliary science goals that complement the technosignature scientific pursuits.

## 4.1 TECHNOSIGNATURE CAPABILITIES AT RADIO, MILLIMETER, AND SUB-MILLIMETER WAVELENGTHS

Considered here are capabilities for both targeted and target-agnostic searches. The targeted searches for technosignatures are analogous to targeted searches for biosignatures, in that one can pose the question of whether there exist atmospheric signatures of life on extrasolar planets (for which there is considerable community effort to measure). By analogy to absolute magnitudes, we consider targeted searches scaled to a distance of 10 pc; Enriquez et al. (2017) provide a more general approach.

At wavelengths of order 1 meter, the number of assets in the United States include the GBT (current longest wavelength of 1.5 m) and a number of small arrays such as Low Frequency All Sky Monitor (LoFASM; University of Texas at Brownsville), which is an array distributed throughout the United States optimized to work at $\lambda$ = 15 m. The Very Large Array (VLA) and Long Wavelength Array (LWA) also both have some capability from 1–10 m.

Internationally, there are a few different arrays working at these long wavelengths, including, but not limited to:

- Low Frequency Array (LoFAR, λ =  3.1–1 m), which is spread throughout western Europe
- Murchison Wide-field Array (MWA, λ = 1–3.1 m) in Australia
- Giant Metrewave Radio Telescope (GMRT, λ = 0.2–10 m) in India
- Hydrogen Epoch of Reionization Array (HERA, λ = 1–2 m) in Australia

Note that none of these instruments have transmit capability.

Moving to the 10 cm regime, the state-of-the-art existing assets are capable of obtaining a signal-to-noise ratio (SNR) of approximately unity in 1 second for transmitters with EIRP of order 100 GW. Examples of telescopes with this capability include the GBT in West Virginia, the VLA in New Mexico, Arecibo Telescope in Puerto Rico, and the GMRT in India.

At wavelengths shorter than 1 cm, the state-of-the-art existing assets are capable of obtaining an SNR of approximately unity in 1 second for transmitters with EIRP of order 100 MW. The notable example is the Atacama Large Millimeter/submillimeter Array (ALMA). The GBT is also capable of receiving at these frequencies at sensitivities similar to ALMA in the overlapping 3 mm range.

For target-agnostic searches, we characterize the searches in terms of the available parameter space, or a Cosmic Haystack. From the perspective of telescope design, there are three parameters that can be controlled by a (radio) telescope designer: the (effective) collecting area of the telescope, its field of view or mapping speed, and the instantaneous bandwidth. (One might also consider the system temperature, except that modern radio receivers are generally near fundamental limits such that improvements in system temperature are marginal relative to the other three parameters.)

In the near term, effective collecting area at the wavelength of interest ranges from 240 m (Arecibo) for limited sky view and frequencies, to roughly 80 m (GBT and VLA) across the 0.2–115 GHz range, and then ALMA at the higher frequency ranges. There are no planned and funded projects exist that will lead to substantial improvements in capability. (In the longer term, the next-generation VLA and the Square Kilometre Array (SKA) both offer the possibility of substantial increases in collecting area, but these projects will likely not bear fruition until the 2030s.)

The instantaneous bandwidth determines the number of signals that could be detected, the character of signals that could be detected (e.g., spread spectrum), or both. Here the state of the art is limited primarily by funding and feed design. The widest feeds on any given can typically only cover 2–3 decades of frequency (e.g., 0.2–0.4 GHz or 2-4 GHz). Wider bandwidth feeds are being developed, which can cover up to 4–5 decades of frequency, but increased bandwidths will come at the cost of sensitivity.

The field of view affects what spatial volume can be sampled instantaneously.

At wavelengths longer than 1 meter, telescopes are generally implemented using elements with naturally large fields of view. Notable exceptions are the GMRT and HERA, but essentially all other telescopes have fields of view of tens of square degrees, up to and including an entire hemisphere.

At wavelengths of order 10 cm, telescopes are almost universally implemented with parabolic elements. The field of view rarely exceeds 1 $deg^2$, but field of view expansion technologies (arrays of feeds or phased arrays) have shown significant recent progress in expanding to as large as of order 10 $deg^2$. The Parkes Radio Telescope, Arecibo Observatory, and the GBT are equipped with multiple feeds, new phased array feeds have been demonstrated on the GBT with a funded implementation project underway for Arecibo, and the Australian SKA Pathfinder (ASKAP) is undergoing commissioning.

At wavelengths shorter than 1 cm, telescopes have narrow fields of view. The GBT is outfitted with a multipixel system that works at 3 mm wavelengths, and the Large Millimetre Telescope (LMT; 50 m dish) has arrays working to smaller frequencies. Field of view expansion technologies are a rich research area with many technologies ready or near ready for deployment on existing telescopes.

# 4.2 TECHNOSIGNATURE CAPABILITIES AT ULTRAVIOLET, OPTICAL, INFRARED (UVOIR) WAVELENGTHS

Similar to radio/mm wavelengths, optical searches are geared primarily in the field of view accessibility, and whether it is a targeted search (single sources) or wide-field searches. Both space-based and ground-based facilities that would offer unique technosignature research fields are highlighted here. Many of these facility-class instruments are not currently used for technosignature research

## 4.2.1 Targeted UVOIR Searches

There are numerous facilities in operation, or soon to be so, which can provide important information for targeted searches for technosignatures. These fall into two categories: 1) looking directly for anomalies that might be due to a technosignature such as a laser pulse or an excess of heat (e.g., Dyson sphere); or 2) identifying candidate stars, such as potential habitable zone Earths, which might become targets at other wavelengths, e.g., searches for radio leakage radiation.

## 4.2.2 Space-based Facilities

Current space-based facilities (Table 3) offer a wealth of archival data that can be searched for transient or unique signatures. It is important to note that search algorithms and analysis for technosignatures are unique from other astrophysics algorithms and processing techniques, and that development in technosignature analysis is highly needed and warranted. It is not simply good enough to state that current data sets or observational programs exists. Similar to radio wavelength, resources are needed to expand in this field.

Table 3 also lists space-based facilities covering a large fraction of the sky, which might reveal unexpected technosignatures or identify solar system interlopers such as `Oumuamua on suggestive trajectories from a nearby star. These missions include all-sky IR surveys such as IRAS, AKARI, WISE, and Near-Earth Object Wide-field Survey Explorer (NEOWISE); Gaia's visible survey, which will identify new planetary systems and refine fundamental stellar properties for over a billion objects; and the Transiting Exoplanet Survey Satellite (TESS) and PLAnetary Transits and Oscillations of stars (PLATO) missions, which will identify nearby transiting systems.

**Table 3. Space-based facilities that would be of use for targeted studies of individual objects. These include missions such as Hubble Space Telescope (HST), James Webb Space Telescope (JWST), and Atmospheric Remote-sensing Infrared Exoplanet Large-survey (ARIEL), as well as archival data sets from Herschel and Spitzer.**

| Facility | Wavelength | Domain |
|---|---|---|
| ***Space-based Assets for Targeted Studies*** | | |
| Active: HST, Spitzer, Stratospheric Observatory for Infrared Astronomy (SOFIA)<br>Archival: Spitzer, Herschel | Visible, near-,mid-,far-IR | Transit spectroscopy, imaging, IR excess searches |
| JWST (2021), ARIEL (2028) | Near-, mid-IR | Transit spectroscopy, imaging |
| ***Space-based Assets for All-Sky Studies*** | | |
| Active: NEOWISE<br>Archival: WISE, Akari, IRAS | Mid-, far-IR | Imaging, spectra, solar system objects |
| Gaia | Visible | Astrometry, time series, spectra, fundamental stellar properties |
| TESS, PLATO (2026 | Visible | Transit detection, time series |

### 4.2.3 Ground-based Facilities

Table 4 highlights ground-based facilities for targets and all-sky searches. Precision radial velocity to find and characterize the closest habitable zone Earth analogs is a critical capability, particularly since transits will only identify a small percentage of the most carefully aligned systems.

**Table 4. Ground-based facilities.**

| Facility | Wavelength | Domain |
|---|---|---|
| ***Ground-based Assets for Targeted Studies*** | | |
| NN-explore Exoplanet Investigations with Doppler spectroscopy (NEID), Keck HIRES, High Accuracy Radial velocity Planet Searcher (HARPS), Habitable Zone Planet Finder (HPF), Lick/APF, other precision radial velocity (PRV) instruments | Visible, NIR | Transit follow-up, identify nearest habitable zone planets, laser pulse searches |
| Keck, Very Large Telescope (VLT), Gemini | Visible, NIR | Follow-up imaging and spectroscopy |
| Next Generation Optical/InfraRed (OIR) Telescopes (2024–2030) | Visible, NIR | Spectra of habitable zone planets orbiting M stars, Follow-up imaging and spectroscopy |
| ***Ground-based Assets for All-sky Studies*** | | |
| Panoramic Survey Telescope and Rapid Response System (Pan-STARRS), Zwicky Transient Facility (ZTF), LSST | Visible | High cadence time series covering entire sky |

# 4.3 SOLAR SYSTEM ARTIFACTS AND INTERLOPERS

Because our exploration of the solar system remains so incomplete (Haqq-Misra and Kopparapu 2012), it remains a potential search space for technosignatures.

Most obvious is the investigation of hypothetical interstellar probes. For instance, the asteroid `Oumuamua is of extrasolar origin (Meech et al. 2017, Bolin et al. 2017, Trilling et al. 2017) and has been the subject of investigation as a potentially artificial object (Enriquez et al. 2018, Harp et al. 2018, Bialy and Loeb 2018). Identification of more such "interlopers" is currently of high interest to planetary science and astronomy, as they provide a unique opportunity to study and potentially even collect samples from extrasolar asteroids and comets. Their study as potential technosignatures thus naturally piggybacks on these studies of interlopers as natural objects.

If such probes exist, it is possible that they might remain in the solar system, either by design or through accidental gravitational capture (Ginsburg, Lingam, and Loeb 2018). Searches for minor planets are naturally also searches for such objects, which might be distinguished by their anomalous orbits (Freitas and Valdes 1980, Freitas 1983, Denisenko and Lipunov 2013), spectra, or colors. If such probes make hard or soft landings on solid surfaces in the solar system, they might be identified through imaging or radar measurements of terrestrial planets and asteroids made for other purposes by NASA assets (Davies and Wagner 2013). Because of the dubious history of unscientific and credulous work in this this area that gained widespread public notoriety (e.g., Carlotto 1997, 1988), care must be taken in approaching, framing, and explaining such searches.

Because the geological, paleontological, and archaeological records on Earth are so incomplete, it is even possible that the Earth itself hosts such artifacts, although, again, this idea is often conflated with unscientific popular imaginings and science fiction stories about alien visitation, and so must be approached carefully.

If technosignatures were discovered in the solar system, it would be worth considering whether their origin might not be interstellar. Specifically, since the Earth is home to the only known species capable of interstellar communication and planetary travel (although both technologies remain in their early development), the Earth remains the only known planet fecund enough to promote technological life, and so it or an early, habitable Mars or Venus could even be the *origin* of such technology (Wright 2017). Gavin and Frank (2018) have suggested that previous episodes of widespread, planet-altering technology on the Earth by putative, now-extinct species (that existed long before humans did) might be identified through paleoclimate investigations using isotopic proxies, land-use analysis, transuranic elements (or fission byproducts), or by searching for artifacts in the geologic record.

More broadly, studies on the historical distribution of intelligent and technological life on Earth (including extinct and extant non-human animal intelligence) inform the terms fi, fc, and L of the Drake equation, and so have direct bearing on searches for technosignatures in the galaxy. Studies of contingency versus convergence and the multiple pathways for intelligence and complexity (e.g., Clayton et al 2017, Blount et al. 2018), inform speculations about the nature and prevalence of intelligence in the universe (e.g., Marino 2015, Morris 2011, Marino and Denning 2012). Empirical and modelling approaches to the durations and dynamics of human societies can inform consideration of L, not just in term of the ranges of values, but in terms of the underlying questions of 'what complex societies do' (Denning 2009; Motesharrei et al. 2014). This work is necessarily interdisciplinary, drawing on evolutionary biology, anthropology, paleontology, planetary science, astronomy, geology, and archaeology, to name a few. In particular, archaeology has long wrestled with the problem of both defining technology and identifying it among natural confounders, problems *in situ* searches for technosignatures in the solar system also face.

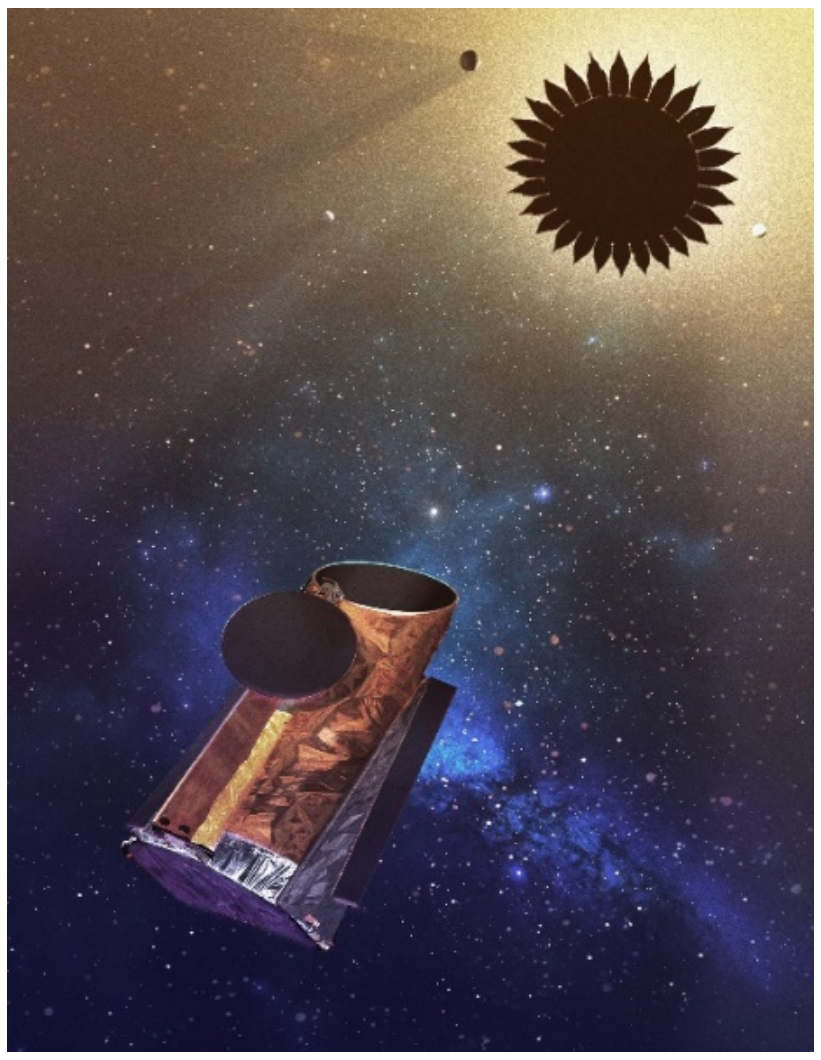

**The HabEx Observatory, one of four mission concepts under study for NASA. Credit: NASA/JPL**

# 4.4 DEVELOPING TARGET LIST FOR TECHNOSIGNATURE RESEARCH

The facilities listed in Tables 3 and 4 can be used to develop a target list. Via transits and precision radial velocity, IR excesses will have identified targets for targeted searches for radio or laser beacons, or radio leakage signatures. In the longer term, studies with possible future mission concepts such as the Habitable Exoplanet Observatory (HabEx) and Large UV/Optical/IR Surveyor (LUVOIR) will identify habitable or even planets with signs of primitive life. These will be prime targets for intensive study.

# 4.5 CONNECTIONS AND POTENTIAL PARTNERSHIPS WITH INDUSTRY

There are three main areas in which partnerships between the private sector and federally funded research in the technosignatures fields can overlap: data storage and access, signal/data processing, and data analysis tools.

In terms of data storage, partnerships with industry are primarily a one-way street, with the majority of the benefit coming to the scientists working within the technosignatures field. The industry partners gain only the prestige/claim of working with scientists and helping them out, and the advertising value, which might be inherent in such a claim. There may be some additional benefits gained by industry, such as scientists providing data for some of the industry tools (e.g., Google Sky), but such uses of science data are fairly limited and specific.

The possibility for partnerships is much higher when considering the data gathering and processing requirements. Focusing on radio wavelengths, the technosignature searches are typically pushing the hardware to their extreme limits. This is true when looking at almost all aspects, from receiver development (sensitive receivers, which maximize the frequency bandwidth and sky coverage) through signal processing (needing extremely high processing speeds to achieve high bandwidth and time/frequency resolution). This need to push to the extreme capabilities of hardware is of significant interest to private industry because it both uncovers flaws and issues with hardware that otherwise may take years to discover and also display a clear path forward for the next generation of instruments. Many of the developments and modifications made to accommodate the needs of high data rate radio astronomy projects, such as the search for technosignatures, are later incorporated into standard within a given industry. This can be seen in the development and use of GPUs, FPGAs, cooled amplifiers, phased array feed systems, and in many other areas.

Finally, in the area of data analysis tools there are many overlaps between the cutting-edge algorithms and techniques developed by scientists working in the technosignatures field and the desire of industry to pull deliberate (and non-intentional) signals from the data. For example, many of the algorithms developed by technosignature scientists working in the radio realm are of direct interest to anyone working to isolate a single (or multiple) radio beacons and pull those from the noise of other signals of non-interest, such as someone looking to isolate the signal from a single satellite or other transmitting device. Similarly, the ability to manipulate large data sets to search for a given signal, which is changing potentially in both frequency and in time is needed for technosignature scientists, other scientists, and anyone who wishes to find and retain a clean signal for later use.

## 4.6 DATA ANALYSIS

Given the latest development in data science and new techniques, such as machine learning and deep learning, new discoveries can potentially already lie in the existing data. Therefore, accessing data specific to SETI and making it available is an area with lots of potential and that can be easily leveraged.

One area where data science and machine learning has been applied is in the biosignatures field. Biologists, chemists, climatologists, and data scientists have been looking at various ways to simulate data when there was no data existing as well as to align various data sets when the data has been disparate or comes from very different fields.

Currently, there have been projects focused on using large data analyses and machine learning in the astrobiology field, in order to simulate data with respect to metabolic genomes and their effects on the atmospheres, or to simulate different potential atmosphere compositions around greenhouse gases. Two such projects were conducted at the NASA/SETI Frontier Development Lab during June–August 2018, an artificial intelligence accelerator for planetary defense problems. These projects simulated more than 300,000 atmospheres using GPU computing as well as computational resources made available by Google. The code base for these simulations was ATMOS, developed at NASA Goddard Space Flight Center. The aim of this project was to couple the types of atmospheres with metabolic network simulated data, in order to understand the coevolutionary effects from micro molecules and genomes to macro atmospheric gases. For the metabolic networks simulations, the teams used code base developed by third parties, such as metabolic network simulations of E. coli genomes (Monte Carlo) from Santa Fe Institute (Eric Libby and Chris Kempes) and ExoGaia agent-based simulation of planetary temperatures, albedos, biomass and atmospheres, developed by Tim Lenton in UK. The teams were successful in coupling the metabolic networks with the ExoGaia planetary temperatures and biomass, but more work needs to be done in order to simulate more genomes and couple them directly with atmospheric gases. The details of these projects can be found in the NASA Frontier Development Lab Technical Memorandum (2018).

# 5 Emerging and Future Opportunities in Technosignature Detection

Svetlana Berdyugina, Natalie Batalha, Rana Adhikari, Amedeo Balbi, Ross Davis, Gabriel G. De la Torre, Kathryn Denning, Duncan Forgan, Adam Frank, Eliot Gillum, Jeff Kuhn, David Grinspoon, Jean-Luc Margot, Sofia Sheikh, Frank Soboczenski, Jill Tarter, Sara Walker, Jason Wright, Shelley Wright

## Introduction

The purpose of this section is to review prospects in technosignature detection, considering opportunities that are currently emerging or being developed, or yet unforeseen. Two key questions are addressed here:

- What new developments would be important for future advances in the technosignature field?
- What role can NASA partnerships with the private sector and philanthropic organizations play in advancing our understanding of the technosignature field?

To understand the future potential of the technosignature field, we first, in Section 5.1, provide an overview of possible technosignature from a theoretical point of view and their cross-disciplinary relations. This helps us then, in Section 5.2, to identify new and possible surveys, instruments, technologies (for both detecting signals on Earth and transmitted by advanced civilizations), and data-mining algorithms. We emphasize that this review aims to summarize a variety of approaches, rather than specific projects. Searches for technosignatures can both supplement and complement other work in astronomy, including searches for biosignatures, and so can proceed both independently of, and in tandem with, other work to detect extraterrestrial life in the universe.

Section 5.3 discusses possible partnerships with NASA that could help advance the technosignature field and create societal impacts. In particular, a public investment in NASA technosignature research provides the opportunity for amplifying NASA's capacity to plan, organize, conduct, and evaluate technosignature research, such that it can overcome challenges and realize significant returns on such an investment. Potential returns on investment include advancing science, expanding commercial applications and jobs, and bolstering public-private partnerships. Ultimately, this can promote the long-term sustainability of technosignature research, whereby the returns on public investment can help spur substantial private investment.

This review is not by any means a complete overview of the subject, but is intended to help stimulate the community in pursuing thorough and extensive research.

# 5.1 UNDERSTANDING TECHNOSIGNATURES FROM THEORETICAL PROSPECTS

There is an inherent need to develop a theoretical basis for technosignatures characterization and evaluation. Much like the work done for biosignatures, theoretical approaches to technosignatures will rely on using principles from physics, chemistry, biology, and social science to articulate avenues through which the evolution of technological energy-harvesting civilizations can leave imprints in signal carriers detectable from the Earth. Ultimately, such a study will lead to designing new instruments dedicated and optimized for technosignatures searches as well as novel approaches in data analysis.

Because (1) the paths of biological cultural and technical evolution elsewhere are inherently difficult, and at some level impossible, to predict and because (2) any society we detect or come into contact with will almost certainly have been in existence far longer than our own, it is difficult to say with any certainty what the observable signatures of such a civilization may be. It is also possible to come across a short-lived outlier with equally unpredictable observable qualities.

The need for specific well-posed research questions may come into conflict with the necessary humility involved in realizing this inherent difficulty of unpredictability. It may well be that, as with finding life, “we'll know it when we see it” could end up being the best guide, and this realization should foster an attitude of searching widely with special attention paid to anomalous observations that do not have initially obvious non-biological, or non-technological, origins. It is also possible that we may come into contact and not be able to fully recognize some technosignatures, which requires developing self-testing analysis systems (e.g., De la Torre and Garcia 2018).

However, such an attitude is not conducive to crafting an observational or theoretical research program. Thus, with these caveats in mind, we can still attempt to derive some general guidelines for the observable characteristics of very long-lived planetary civilizations and possible short-lived “bursters.” Beyond specific predictions of temporal developments of technological capabilities and signatures, we can ask: What are the likely characteristics of very long-lived planetary societies, and of planets that have been modified by long-term co-evolution of technology with planetary physical and biogeochemical cycles? As to short-lived civilizations, those may be much “louder,” and thus prominently observable, while their planets' characteristics may bear signatures of short-term, intense, and unsustainable coevolution with technology. There has been some theorizing about postbiological intelligences, which need not be planetary, and could in theory dominate the universe. Those pose their own modeling challenges, which are not addressed in this review.

One guiding principle that may be useful is considering the overlap of two questions. 1) What features of a civilization as a physical phenomenon can't be avoided? This question can take on a thermodynamic or information-theoretic character: what tracers can and cannot be avoided in the process of civilization building? 2) Which of these tracers can be detected? Which kinds of signals will be robust enough to be separated from noise and be clear enough to not be confused with natural processes? This schema of “what can't be avoided/what can be detected” can help guide new avenues of research in technosignatures, especially in the exoplanet era.

This section distinguishes among communication, atmospheric, structural, and global (planet-scale) technosignatures, because their detection requires different observational approaches, followed by a brief discussion of possible time evolution of technosignatures and their cross-disciplinary significance. In many cases, synthesizing observations of technosignatures and modeling strategies of observations are useful aids to future survey designs.

## 5.1.1 Communication Technosignatures

### 5.1.1.1 Communication as a Technosignature

Communication is an information-processing technique. All terrestrial life forms communicate with other life forms and, in thermodynamic terms, also with the environment. In fact, "life cannot not communicate," according to Watzlawick's famous axiom. Hence, early ideas on how to search for extraterrestrial intelligence were formulated using and foreseeing examples of human communication technologies based on radio, TV, and lasers (Cocconi and Morrison 1959, Dyson 1960, Schwartz and Townes 1961). The current state of these communication technoignatures is reviewed in other chapters of this NASA coordinated effort to understand prospects of technosignature research. It seems logical that these "classical" radio and optical/IR SETI programs continue until the solar neighborhood is carefully scrutinized for possible signals, including broadcasting unintentional and intentional, Arecibo-like messages. Section 5.2 provides a brief commentary on approaches to improve the efficiency of such SETI surveys.

**On this front, the medium-term future of radio telescope development includes the deployment of the full Square Kilometre Array, which will be the first telescope with the sensitivity to detect Earth 2020-levels of radio "leakage" at interstellar distances. This is an important milestone in radio technosignature work, since it marks the first time that humanity will have the capability, in principle, to detect a clone of itself on a planet around another star without any assumptions about coordination or transmission intentions. It then stands to reason that any technological species with stronger radio transmission leakage could be ruled around the nearest stars.**

### 5.1.1.2 "Bottled" Messages and Alien Artifacts

With current technology, space-based alien artifacts, possibly bearing messages on board similar to Pioneer's and Voyager's "bottled" messages, could be detected within the solar system, using reflected light polarimetry, since artificial materials appear in polarized light very different from natural ones (e.g., Kissell 1974; see also Section 5.2.2). Similarly, alien artifacts on surfaces of solar system objects could be detected. Rose and Wright (2004) suggested that long, information-rich messages could be inscribed at near-atomic densities into artefacts and sent slowly and redundantly to a target with

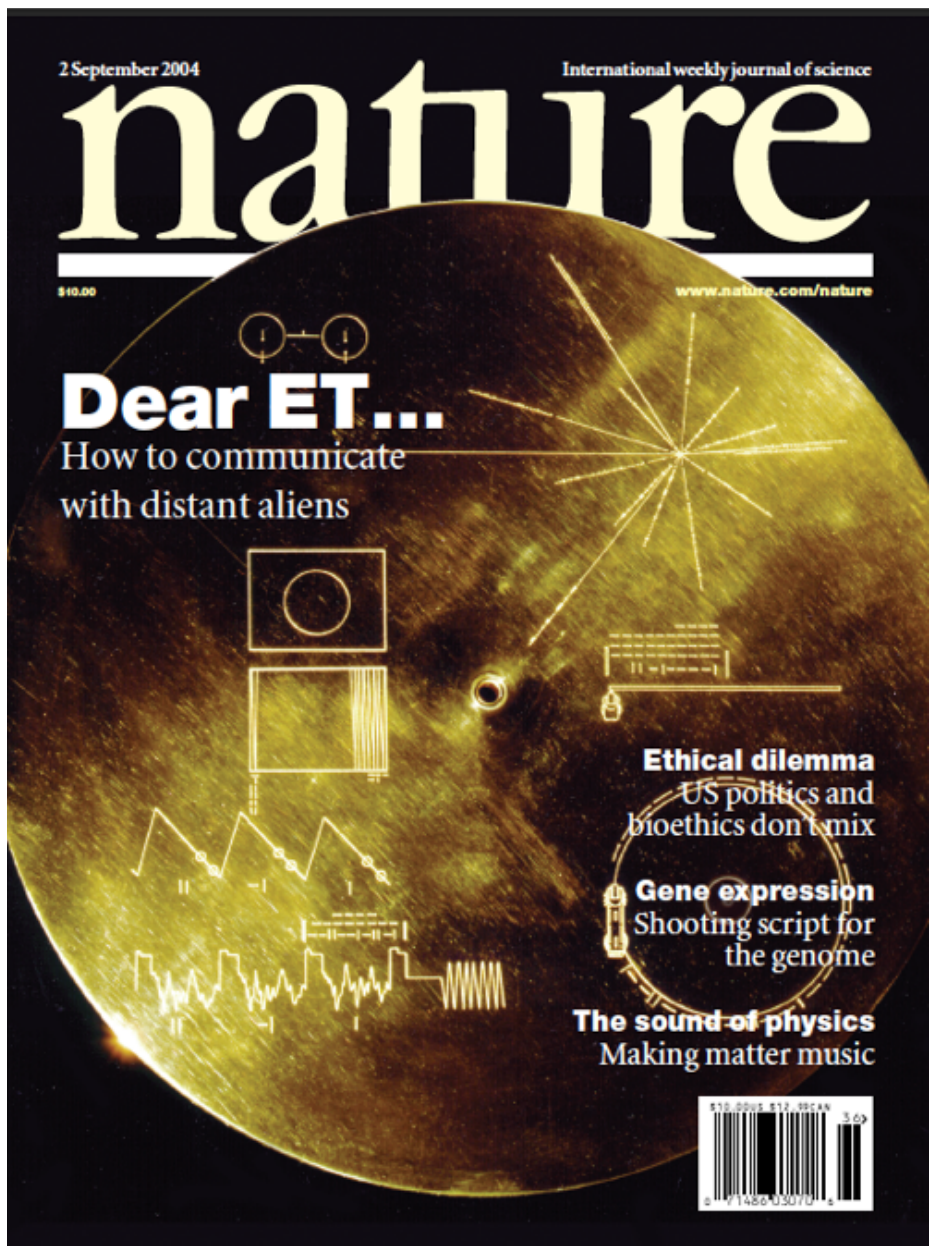

Credit: *Nature* International Weekly Journal of Science.

greater energy efficiency per bit than using electromagnetic messaging.

#### 5.1.1.3 ALIEN ARTIFICIAL INTELLIGENCE

Artificial intelligence (AI) offers great opportunities to process large amounts of data quickly and accurately. It employs different methods, including artificial, convolutional neural networks, pattern recognition, machine learning, fuzzy logic, etc. AI is currently being used in several scientific disciplines including engineering, medicine, computing, robotics and economics. Its association with machines and robots offers new boundaries to help in space exploration.

Within communication domain, current studies span between centralized AI systems and distributed AI. As we learn to communicate with AI for our own benefits, we may become better in recognizing alien AI signals. Moreover, a possibility exists that in the near future, some type of communication may happen between an Earth-based AI supported system and an extrasolar form of AI.

However, AI utilization in the search for technosignatures may first be outside this communication aspect, helping in big data analysis, pattern recognition, and discrimination (relevant to Section 5.2.6). It has not yet been clarified if this AI will be free of perceptual and experiential bias in opposition to its human counterpart. For this reason, the development of AI systems to support the search for technosignatures and data analysis, mining, and processing may represent a key aspect in the near future.

### 5.1.2 Atmospheric Technosignatures

One of the most obviously artificial ways humanity has altered the Earth's observable characteristics is via atmospheric pollution. Regardless of whether artificial constituents of an atmosphere are present as unwanted byproducts of industry (literal “pollution") or as a deliberate form of geoengineering, the detection of such compounds may be one of the strongest and most ambiguous technosignatures of a planet-bound species.

Schneider et al. (2010) and Lin et al. (2014) explored this possibility and specifically addressed CFCs as an example. Lin et al. (2014) identified several industrial “pollutants,” focusing on CF4 and CCl3F, to show that such technosignatures could be detectable with less than 2 days of integration time with JWST. However, the Lin et al. (2014) model assumes a U.S. standard atmosphere to represent temperature, pressure, and composition of the planet's climate. CFCs and other organofluorine compounds are potent greenhouse gases, and so the presence of any so-called industrial pollutants will inadvertently cause warming and change the mean climate state. Thus, any synthetic spectra describing spectral technosignatures must also account for any associated greenhouse warming (or cooling). A next generation study would include full climate modeling for greenhouse active artificial atmospheric constituents (AACs) and include a range of planetary conditions beyond those of Earth. This would include planets in a range of positions within the habitable zone and using state-of-art atmospheric retrieval codes to produce synthetic spectra. Stevens et al. (2016) demonstrated that there are several short-lived atmospheric technosignatures that might be detectable during the last stages of a technologically induced planetary cataclysm. Obviously, such pollutants are to be in chemical disequilibrium with other, “natural” atmospheric constituents.

## 5.1.3 Structural Technosignatures

### 5.1.3.1 Artificial Megastructures

Berdyugina and Kuhn (2017) have shown that artificial megastructures can be spatially and spectrally resolved through inversions of planetary reflected light curves acquired in multiple passbands. In particular, subcontinent-size highly reflective artificial structures (to reject unwanted stellar light into space) and highly absorptive, photovoltaic-like structures (energy generators) in the near-planetary space above clouds can be resolved and recognized thanks to their high contrast to the natural environment and a possible regular substructure or “unnatural” shape. Modeling shows that such large, subcontinental-size megastructures can only be detected if they are spatially resolved using an inversion technique, so that their relatively undiluted spectra and polarization can be measured and constituent materials are identified. If megastructures are spatially unresolved, they may remain undetected as their spectral and polarimetric signatures are strongly diluted and reduced by global environmental signals.

### 5.1.3.2 Heat Islands

Kuhn and Berdyugina (2015) proposed that Kardashev Type I civilizations, like ours, confined to energy sources on the planetary scale will be ultimately forced to use exclusively photonic energy provided by the stellar radiation to slow down an unavoidable global warming, as a consequence of the second law of thermodynamics. Assuming that social aspects of terrestrial life is a universal feature, civilization development will be clustered in favorable geographical areas where waste heat is constantly dumped into the environment raising its temperature to remotely detectable levels. Such “heat islands” are well identified with large cities in infrared Earth images taken from space. They have been proposed to be a prominent technosignature for civilizations only slightly more advanced than ours (consuming ~50 times more energy). A network of alien civilization heat islands may be resolved on the planetary surface and distinguished from environmental heat sources (e.g., volcanoes) through differential infrared measurements, e.g., at 5 μm and 10 μm.

### 5.1.3.3 Artificial Illumination

Schneider et al. (2010) and Loeb and Turner (2012) both suggested that artificial light sources might be detectable on the night sides of exoplanets with near-future technology, by analogy with human nighttime city illumination. Kipping and Teachey (2016) explored the possibility that light might be deliberately transmitted in the anti-stellar direction and be detectable during exoplanetary transit.

### 5.1.3.4 Constraints from Complexity Theory

Complex systems theory, particularly scaling laws, as well as the theory to generate them can place constraints on expectations for alien technological architecture. Examples include structure of communication networks, cities, etc. Such scaling laws of cities coupled with imaging data of planetary surfaces using inversions of planetary reflected light curves (Berdyugina and Kuhn 2017) can provide information on alien sociology.

## 5.1.4 Planet-scale Technosignatures

### 5.1.4.1 Energy Dissipation as a Technosignature

Frank et al. (2018) proposed a classification system for planets based on the dissipation of free energy generated within the coupled planetary systems. It was demonstrated one can expect a rise in free energy dissipation when going from lifeless worlds with atmospheres through worlds with robust biospheres and on to planets with long-term (sustainable) technological civilizations. The creation of free-energy gradients from emerging complexity within the planetary systems, in particular a biosphere, falls into the category of "what can't be avoided," discussed above. Theoretical studies are needed to articulate what kind of technosignature this might be imprinted in the light from a planet with a long-term planetary scale civilization. We note that this kind of work has been carried out for biosignatures, in terms of atmospheric disequilibrium, and shows promising results (Krissansen-Totton et al. 2018).

### 5.1.4.2 Climate Stability or Instability as an Effect of Civilization

General qualities of planets that have been inhabited over long timescales by technological civilizations may have some characteristics of stable equilibrium. One might reasonably postulate that such a civilization would value environmental stability and would thus attempt to interrupt any natural "Milankovitch cycles" or other periodic or episodic cycles of dramatic climate change. Thus, (for example) if planets are observed that appear to have been modified in such a way as to dampen or eliminate such climate instabilities, this could be regarded as a potential technosignature. As has been demonstrated by Haqq-Misra (2014), Earth's Milankovitch cycle of glaciations may likely have already been altered by our inadvertent interference in Earth's climate system. This makes the intentional disruption of such cycles by even a slightly more "advanced" civilization (i.e., for this purpose, one with a greater level of planetary-scale long-term coherent intentional application of technology) seem quite plausible. In particular, global warming could be a planet-scale thermodynamic technosignature (Kuhn and Berdyugina 2015). On the other hand, exacerbation of cycles may be seen as a civilization lurch along dysfunctionally with incomplete cycles of social collapse and regeneration for a long time and on a planetary scale.

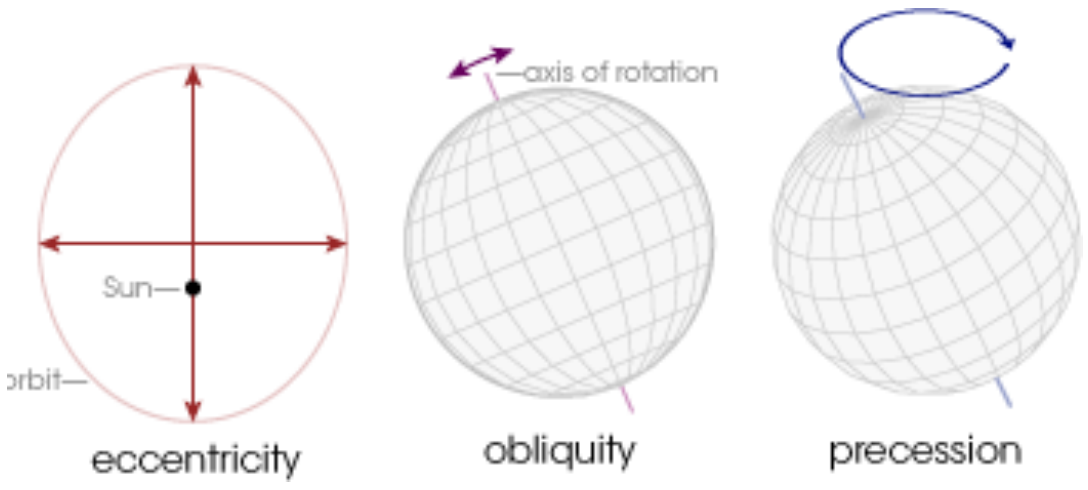

**Milankovitch cycles. Credit: Robert Simmon, NASA GSFC.**

### 5.1.4.3 Geoengineering

It is conceivable that the signs of planetary engineering may extend to more than one planet in a multi-planet system. One can imagine that in a system with several habitable, or potentially inhabitable, planets, a society may desire to engineer nearby planets to have similar climates to accommodate a biosphere extending to multiple nearby inhabited worlds. Thus, one possible technosignature may be the observation of several planets in one system with suspiciously similar climates (that is, those that do not likely seem to be the result of coincidental separate evolutionary paths), which may represent such multi-planet scale tinkering. This tinkering may even extend to the belts of debris present in these systems—mining asteroids may provide valuable resources for large-scale engineering, and this can yield its own technosignature (Forgan and Elvis 2011).

Lingam and Loeb (2017) explored the spectroscopic signature of major geoengineering efforts associated with planet-scale stellar energy collection, such as surface silicon photovoltaics and chlorophyll-containing material, finding that both produce characteristic, detectable spectral edges. While such activity would be obviously detrimental to the native biosphere of such a planet, it might be a more benign endeavor on an otherwise dead world (for instance, in the far future humanity might contemplate such a project on the Moon or Mercury). Such technological spectral edges may not be easy to distinguish from biological photosynthetic edges (Berdyugina et al. 2016), as the former are developed to mimic the latter.

Following further the potential desire for environmental and climate stability as a value that could plausibly be held by otherwise inscrutable or unpredictable exocivilizations, we can imagine that if a planet was inhabited over an extremely long time-span, approaching relevant timescales for stellar evolution, that planetary engineering might be undertaken to avoid the inevitable runaway greenhouse that will befall every Earth-type inhabited planet in a stellar habitable zone as stellar main sequence evolution progresses and increased stellar luminosity gradually sweeps the habitable zone outward. Thus, planets that appear to be anomalously preserved in a climate state that represents the climate balance of some earlier epoch of stellar evolution might plausibly have been engineered to preserve a past climate state to which that planet's biosphere had become well adapted.

This sample list of possible planetary evolution technosignatures is meant to be illustrative, not exhaustive. Indeed, as described above, there are good reasons to believe an exhaustive list is not possible. Nonetheless, a research program that investigates likely physical planetary limits that would be encountered by any long-lived civilization, and, considering sample sets of plausible goals and values for such a civilization (such as environmental stability, or maximizing biomass, etc.), likely evolutionary states can be modeled. This may suggest fruitful observational programs as well as better preparing us to interpret the anomalous and puzzling signatures that may ultimately be the first indications that such an entity is being observed.

### 5.1.5 Time-Evolution of Technosignatures

The quantity and type of technosignature existing at a certain epoch in the universe is probably not stationary, due to a number of astrophysical, biological, and even sociological factors that depend on time. Modeling the temporal distribution of technosignatures over cosmic history is crucial for a number of reasons. Better modeling would result in:

- More sophisticated estimates of the number of potential signals we might intercept from our location in space and time;
- Indication of the most probable kind of technological species producing the signal (e.g., in terms of energy capability, signal strength, signal content, and so on); and
- Help in making predictions of the signal to compare with data, which would be especially useful when trying to interpret anomalies.

All of these would help in planning optimal observational strategies. A theoretical framework of this sort would also give us a useful benchmark to interpret possible outliers (such as candidate signals from civilizations that are in a stage of technological development similar to ours).

A simple parameterization of evolutionary effect might involve two timescales (Balbi 2018): the time of appearance of species capable of producing technosignature and their longevity (broadly defined as the duration of technosignature, which is independent on the actual survival of the species). The statistical properties of such timescales would define a probability distribution of technological species, which can be used to produce simulations. Based on this parameterization, the number of technosignatures that can be detected by a survey observing a given volume of space centered around our location can be estimated by simple causal considerations. Both timescales depend on a complex interaction of factors. The time of appearance of technological species would depend, for example, on the habitability history of the universe (Lineweaver et al. 2004; Loeb et al. 2016), on the probability of abiogenesis on habitable planets, and so on. The longevity of a technosignature is arguably more difficult to model and less studied so far, because it depends not only on astrophysical or biological factors (e.g., how long a planet can in principle remain habitable, or on the characteristic scales of biological evolution for complex life) but also on sociological issues, as well as on intentionality (e.g., whether technological growth is sustainable in the long run, whether a species is willing to communicate, whether it builds structures or beacons that can survive its extinction and even that of the planetary environment, etc.).

Another parameterization might involve the notion of a transition probability to extreme longevity. In the “bottleneck” model described qualitatively by Grinspoon (2016), there may be a bifurcation in lifetimes between those civilizations whose lifetimes are limited by the existential threats experienced by young technological civilizations (represented, at least in part, by the many known potential existential threats to our own civilization  as summarized, e.g., by Rees (2003), Wilson (2002), Avin et al. 2018, and those who have developed understanding to the point where such threats are transcended and cosmological-length lifetimes are achieved. In such a scenario, the Drake equation, which assumes a fixed value of $L$, and therefore assumes a steady-state number of technical civilizations, is not valid, and a time-dependent equation, in which extremely long-lived civilizations (quasi-immortal in Grinspoon 2016’s lexicon) will accumulate over time, will apply. Quantitative modeling of these scenarios can help to refine the potential observational search space.

On the other hand, our history demonstrates that civilizations collapse but incompletely and secondary civilizations arise. If climate changes or resources overshoot, similarly, some parts of a planet will have the capacity for local mitigation while others don't. Hence, decimation and extirpation is conceivable rather than complete extinction. Also, high technology could well survive overall civilizational lows. So, a complete bifurcation between “die-young-and-not-too-bright” and “live-forever-in-infinite-wisdom” may not be obvious.

## 5.1.6 Activities in Disciplines Relevant to Technosignatures

### 5.1.6.1 Linguistics and Nonhuman Communication

The detection of any technosignature or biosignature will require continual engagement with the public (not just a single announcement). A signal with semantic content will also immediately require sustained efforts at decipherment/interpretation, and inference about the sender(s) (Elliott and Baxter 2012). Work in computational linguistics (Elliott 2009, 2011) and nonhuman communication on Earth (e.g., Herzing et al. 2018, Kershenbaum et al. 2018, Berea 2018, Doyle 2011, Elliott 2015) and potential semiotic universals (Deacon 2018) is already exploring this, and even minor funding could substantially accelerate progress. Crucially, this research capacity should be built and maintained before a detection, i.e., as an integral component of an ongoing technosignature/biosignature search program. It cannot realistically be manifested out of the blue on an as-needed basis.

In contrast to standard (traditional) decipherment techniques, complex bootstrapping of discovered phenomena, against predominantly unsupervised learning and decipherment algorithms are required for such analysis, to discover, build and assign structural (syntactic) components, across a multilayered affinity matrix, during post-detection analytics (Elliott 2012, Elliott and Baxter 2012): phases crucial for structure categorization to build towards where syntax meets semantics. It is also essential that the framework for post-detection frames itself within an efficient 'handshake' to ongoing detection capabilities and mechanisms for dissemination, as the process unfolds and progresses (Elliott 2008).

**The more we know about what is universally possible and impossible (i.e., what are the physical boundaries of communication?), the better we can frame and prepare decoding and interpretation protocols. The advances in disciplines such as linguistics, animal communication, information theory (Shannon 1948), cryptography, should come together in an integrated way for understanding what is universally possible and impossible in terms of communication.**

#### 5.1.6.2 ANTHROPOLOGY AND ARCHAEOLOGY

The theory, principles, and methods used by archaeologists to search for hidden human technosignatures (artifacts, archaeological sites, and anthropogenic landscape modification) on Earth are continually evolving and should be a source of innovation for solar system extraterrestrial technosignature searches. For close-range solar system searches, archaeological methods may apply, including ground-penetrating radar and magnetometry. Remote-sensing methods such as unmanned aerial vehicle (UAV) photography and Light Detection and Ranging (LiDAR), thermal and infrared imaging, have been adapted and used within archaeology for locating and mapping material culture / archaeological sites on Earth, including subsurface to about 1 m depth.

Anthropology has multiple roles to play in the future development of technosignature theory and modeling. It can be integrated into technosignature/biosignature searches in three ways:

1. As a data-rich evolutionary-historical science devoted to the only species we know of that has produced a technological civilization capable of large-scale engineering and communicating at interstellar distances;

2. As a set of methods and perspectives that can reflexively refine our ability to step outside familiar mindsets to confront alien ways of being; and

3. As a way of thinking about what it means to be human in a complex environment including myriad other life forms (Denning 2018).

First, the archaeological record is a goldmine for building theory, if used to examine the emergence and sustainability of complexity from a cross-cultural comparative perspective (rather than in a unilineal evolutionary narrative, as is the tradition within SETI). It can speak in unique ways to the emergence and intensification of complexity, the relative roles of contingency and convergence, the sustainability of complexity, and variable ecological footprints (Denning 2009). Recent big data initiatives in archaeology, particularly the Seshat Global History Databank,[3] are finally collecting disparate global archaeology / history data into a single database to allow scientific study of questions about the evolution of civilization on Earth. There are already some surprises (Turchin et al. 2018) and more will surely come quickly.

Second, previous models of human social evolution have tended to ignore the reality that we coevolved biologically and socially with other species (at both the micro and macro scales). New modes of anthropological and archaeological thought address this (Denning 2018). For the purposes of technosignature theory-building, we need to integrate this recognition into holistic, planetary models of biological, social and technological complexity.

Third, refined search technologies and methods are certainly essential for next-generation technosignature searches, but we can further develop our intellectual toolkit as well. Examining recent human failures and successes in uncovering and recognizing evidence of unfamiliar intelligence, past and present, in our own world, will be instructive (Denning 2018).

Finally, both archaeological theory and methods should be employed in any searches on Earth for nonhuman technosignatures.

### 5.1.6.3 NEUROSCIENCE AND THE EVOLUTION OF INTELLIGENCE

The emergence of intelligence is a factor in the Drake equation, but the reasons for studying it go well beyond the estimation of this factor. Assumptions about it have been woven into technosignature searches for decades, and there is a substantial task ahead in integrating the current and future best theory and data from bioscience focused on Earth into efforts to locate life elsewhere. This area of biology is accelerating rapidly: questions about the origin and evolution of intelligence on Earth are now more technically, methodologically, philosophically, and scientifically tractable than ever before. The Intelligence in Astrobiology workshops (2012) and Virtual Resource Center on Intelligence in Astrobiology[4] (2009–13; Marino and Denning et al.) are examples of NASA Astrobiology Institute (NAI) funded initiatives working on this integration. Recent papers by Marino (2015), Berns (2018) and McShea (2018) address the state of the field as applicable to technosignatures searches, and flag ways forward.

Neuroscience, including cellular and physiological, cognition, and behavioral aspects, is also currently experiencing a great advance. With the help of neuroimaging techniques we better understand now the nervous system and how mind works. These advances are even being translated into modern computing methods and AI. This rapidly growing discipline is going to be of great help in searching for technosignatures, and also a posteriori in the event of finding a signal or any form of communication.

[3] http://seshatdatabank.info/

[4] www.yorku.ca/astrobio/intelligence/index.html

Once we fully understand how intelligence has evolved, and how it has become cultural, and even technological, in multiple species on Earth, we will know what, if anything, might be universal about that process. If we deeply understand the evolution of intelligence (and consciousness) on Earth, we will be much better situated to predict the possibilities for intelligent life elsewhere, or to comprehend any extraterrestrial intelligence or complex life, should it be found.

## 5.2 NEW APPROACHES TO DETECTING TECHNOSIGNATURES

Commensal observations, during which the search for technosignature is carried out simultaneously with other astronomical observations using general astronomical facilities, are proven to be interesting but so far have not led to a detection. They definitely offer opportunities for maximizing the return on investment of telescope resources, but it becomes more evident that surveys and instruments are to be designed and optimized for the needs of the technosignatures discussed above.

### 5.2.1 New Surveys with Current Instruments

There is a number of optical/IR facilities that have contributed or will contribute to the search for technosignatures via commensal observations. Kepler, K2, and TESS provide continuous photometric data, which are mined for transiting megastructures. Cherenkov array detectors search for serendipitous laser communication signals and other unusual transient optical/IR events. Similarly, arrays of solar heliostats can be used during the night to increase the non-imaging coverage of the sky Covault (2001). With the new arrays coming online, we can approach a desired “all-sky & all-the-time” observations scheme, similar to technosignature searches in the radio. High-resolution spectrographs on large, 8–10 m class telescopes are employed to search for spectral signatures of lasers communication. JWST (2021) may be able to provide upper limits on atmospheric technosignatures on nearby exoplanets. Current and planned surveys with these instruments are discussed in accompanying reviews. Needless to say, such surveys are to be continued and expanded as new facilities come online.

In the radio part of the spectrum, much work is needed to sample a larger fraction of the Cosmic Haystack (Tarter et al. 2010), also known as search volume. The dimensions of the search volume (e.g., Drake 1984, Tarter et al. 2010) normally include sensitivity, sky coverage, frequency coverage, and time. Both targeted surveys and broad surveys are required to examine all dimensions of the haystack. Targeted surveys are necessary to sample regions at high sensitivity, whereas broad surveys are ideal to maximize sky coverage. In both cases, efforts should be made to maximize the frequency coverage and time span of observations (“all frequencies, all the time”). Sensitivity is a particularly important dimension of the Cosmic Haystack because the spatial volume from which transmitting technological species can be detected scales as the minimum detectable flux to the (-3/2) power. Existing instruments (e.g., Green Bank Telescope, Arecibo Observatory, NASA/Jet Propulsion Laboratory Deep Space Network, Parkes Radio Telescope, the Allen Telescope Array, Sardinia Radio Telescope, Five-hundred-meter Aperture Spherical radio Telescope, MeerKAT, etc.) offer significant opportunities for sampling a larger fraction of the search volume in the next decade. With the possible exception of telescope arrays with beamforming capability, there is a substantial penalty associated with commensal observations: the primary observer dictates the pointing and cadence of observations, which prevents the secondary observer from using an optimal cadence (e.g., for RFI mitigation) and/or re-observation strategy (e.g., following the near-real-time detection of interesting candidate signals). These drawbacks emphasize the need for telescope time dedicated to the search for technosignatures.

### 5.2.2 New Instruments: Optical and IR

Considering the scope of technosignatures discussed in Section 5.1, technosignature detection in optical/IR requires:

- High-contrast, large-aperture imaging systems capable of direct imaging of Earth-size planets around G-, K- and M-class stars;
- High- and medium- resolution spectroscopy in optical/IR for detecting atmospheric and optical/IR communication technosignatures;
- Optical photometry and polarimetry for detecting structural technosignatures in transits and reflected light; and
- Infrared photometry for detecting thermodynamic technosignatures and climate change effects.

These instruments are to be highly efficient in harvesting optical/IR photons from Earth-size exoplanets.

### 5.2.3 High-contrast Imaging Systems

The integration time needed at a fixed SNR in direct exoplanet photometry scales faster than $D^{-4}$ (where D is the telescope diameter) and depends critically on exquisite wavefront control (reaching several nanometers) to realize high contrast. Therefore, Extremely Large Telescopes (ELTs) with high-contrast capabilities (better than $10^{-6}$ at small angular separations) are needed to search for technosignatures and biosignatures on exoplanets.

Currently planned ELTs are optimized for relatively wide-field general astronomy and will not necessarily address many exoplanetary science goals, including searches for technosignatures. A dedicated instrument that is capable of high-contrast direct imaging of Earth-size exoplanets in the optical and infrared would enable more sensitive technosignature searches in the optical and near-infrared. To achieve the high contrast with a large enough aperture, a hybrid interferometric-coronagraphic telescope concept was recently proposed (Kuhn et al. 2018, Moretto et al. 2018). In contrast to other ELTs, such a system has a narrow field of view (FOV) and combines interferometric concepts with "floppy" subaperture optics, to dramatically decrease moving mass and cost per $m^2$ of light collecting aperture while realizing extremely high-photometric dynamic range. These

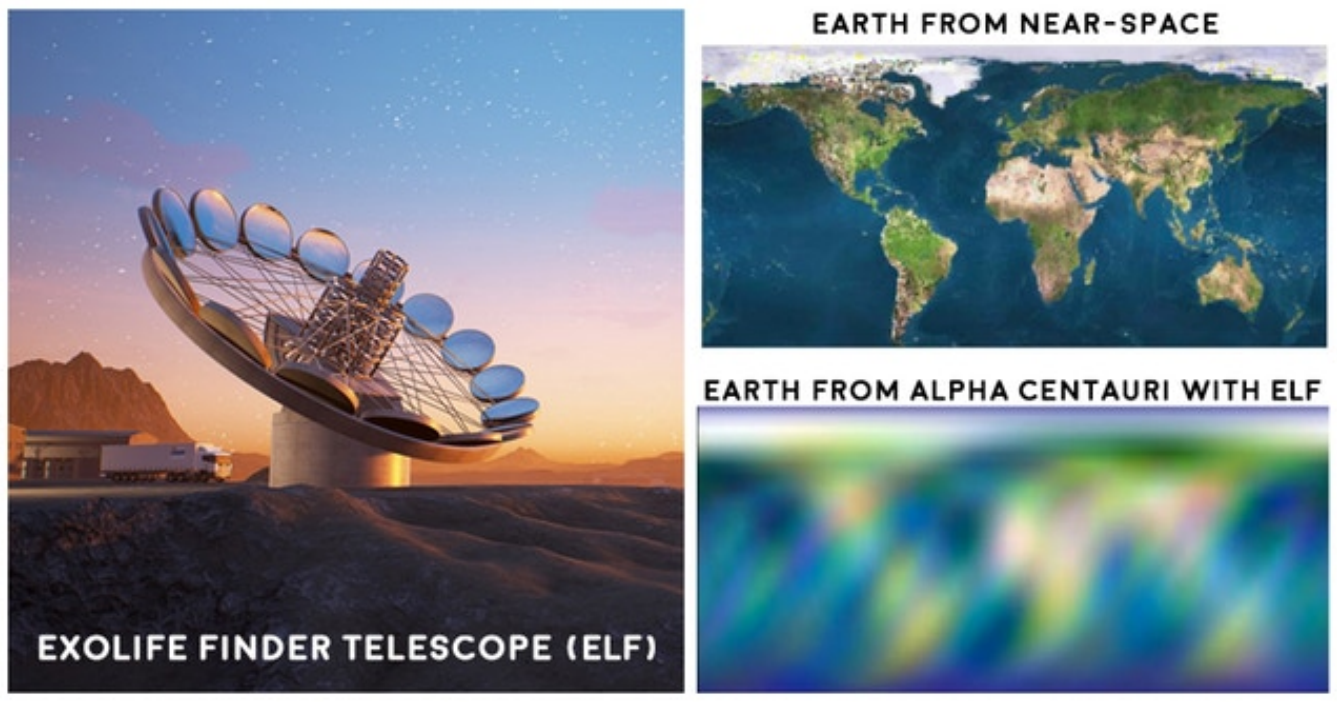

**Illustration of ELF. Credit: PLANETS Foundation.**

advantages make comparable-aperture space-based systems non-competitive in both cost and time to build. One of the realizations of such a system is Exo-Life Finder (ELF), which consists of 9 to 25 large (4–8 m) off-axis telescopes assembled on a common pointing structure. The primary mirror segments have identical off-axis parabolic shapes, and are served by corresponding adaptive secondary mirrors, each creating a diffraction-limited image with high-accuracy wavefront control. The synthesized aperture created by properly phasing the ELF segments can produce a $10^{-7}$ contrast dark spot that can be moved within the FOV by modifying segment phases. Such a high-contrast, large-aperture system is specifically designed to be sensitive to exolife (biosignatures) and exo-civilizations (technosignatures) on many nearby exoplanets (Berdyugina et al. 2018a,b).

#### 5.2.3.1 Spectroscopy

High- and medium- resolution spectroscopy in the optical and IR up to 25 μm has been used for characterizing exoplanets since more than a decade. Such spectrographs are planned for the next-generation ELTs as well. They are not designed specifically for technosignature searches, but standard techniques used for characterizing exoplanetary atmospheres are well-suited to search for technosignatures.

#### 5.2.3.2 Time-series photometry

Direct imaging of Earth-size planets will allow acquiring high SNR continuous reflected light curves at 0.3–5 μm. Inversions of such light curves in various spectral bands will produce detailed exoplanet albedo maps, including possible artificial megastructures on the surface and in the near-exoplanetary space (Berdyugina and Kuhn 2017). Additional IR measurements at 10 μm will allow a search for civilization heat islands (Kuhn and Berdyugina 2015). Also, various atmospheric technosignature (pollutants) could be detected in the optical and IR.

#### 5.2.3.3 Time-series Polarimetry

Polarimetry helps increase the contrast of directly imaged exoplanets (e.g., Langlois et al. 2014), characterize albedo of exoplanets (Berdyugina et al. 2007, 2011), and can reveal microscopic properties (composition and texture) of various surfaces (e.g., Berdyugina et al. 2016). This can further be exploited to identify artificial structures within natural environment, similar to searches of various manmade objects within a rocky or icy terrain (e.g., Snik et al. 2014). Hence, technological artefacts, such as free-floating spacecraft or satellites, may be identified by their anomalous polarization characteristics. For example, reflection of unpolarized sunlight from a spacecraft can be specular or diffuse, depending on whether its surface smooth or rough, dielectric or metallic, or if it is of unnatural geometric shape (Kissell 1974). A combination of surface phase shifts and polarization sensitive absorption and reflection can produce elliptical or linear polarization and provide a diagnostic of the artificial nature of the object. Taken to extreme, technological mega-structures on the scale of a solar system, such as a Dyson sphere, may have reflections and glints that are highly polarized. Time-series polarimetric measurements at various phase angles and inversions of polarized light curves (Fluri and Berdyugina 2010) are needed to reveal such glints.

### 5.2.4 New Instruments: Radio

Types of instrumentation/approaches needed to accelerate technosignature detection in the radio:

- Multi-channel spectrometers (fast Fourier transform (FFT)), to cover a broader range of frequencies;
- Wide field of view, to efficiently catch transient events;
- Machine learning, to analyze data quickly and in novel ways;
- Ring buffers for very large, real-time data sets, to develop ability to dump data from ring buffers to permanent storage on the basis of real-time analysis triggers; and
- Much more massive multipixel arrays on single dishes, next-generation Very Large Array (ngVLA), wideband feeds, and improved signal processing hardware and algorithms.

Detecting signals of different modulations (narrowband, broadband, pulses, signals with irregular structure, etc.) requires also theoretical work on the maximum drift rates that should be searched by our algorithms when looking for linear features, as well as application of convolutional neural networks (such as that used for fast radio bursts (FRBs) in Zhang et al. 2018) and cross-correlation functions. This ties into the machine learning aspect, but it is broader and has applications outside SETI.

### 5.2.5 New Instruments: Non-EM

Traditional EM SETI often involves guessing the wavelength at which a technosignature (such as a deliberate, communicative beacon) might be discovered, and is often driven by the sorts of radiation humanity generates and can detect. As new, non-EM detectors come on line, it is possible to extend the search for technosignatures into these new domains, where technosignatures might exist and compete against very low backgrounds.

In particular, if extraterrestrial technology involves the manipulation of amounts of energy or matter many orders of magnitude beyond humanity's current capabilities, that may manifest in a variety of non-electromagnetic signatures detectable across astronomical distances.

#### 5.2.5.1 Gravitational Waves (GWs)

Gravitational waves can be used for communication over sub-galactic distances. GWs have several advantages over EM waves, and one major disadvantage.

> **Advantages:**
>
> GW detectors measure field, not power, so the signal strength decays as $1/r$ rather than $1/r^2$.
>
> GWs are not scattered by interstellar material or stars. They pass through all but the most dense galactic environments without scattering.

> The GW sky is 'dark' in the audio band. All of the GW energy in the universe from astrophysical sources is below 1 Hz. Thus, beamed GWs have no background to contend with.
>
> GWs may be considered to be an advanced technology in the sense that long-range communication is possible without less-technologically capable observers even knowing the signal exists.
>
> **Disadvantage:**
>
> GWs require an enormous amount of energy to make a detectable signal, at least to Earth-2018 technology.

Since the first detection of astrophysical GWs in 2015, there have been several other detections—all from compact binary mergers. The most sensitive instruments for detection of GWs are the Laser Interferometer Gravitational-Wave Observatory (LIGO) detectors. Over the next 20 years, it is envisioned that the strain sensitivity of these detectors (10–10,000 Hz) will increase by 100x (100x in sensitivity is ~100x in distance), and will be joined by less sensitive space-detectors (0.0001–1 Hz). The prospect for GW detection in the 1–1,000 MHz band is not promising.

Using Earth-2018 technology, and ~$1B, it could be possible to beam GWs across the distance of the Earth. In order to make a signal detectable with Earth-2100 technology, from a distance of a few thousand light years, would require a very large array of mass quadrupoles. A trillion such quadrupoles with a combined mass of $10^{24}$ kg (~1 earth mass), would still be beyond the sensitivity range of Earth-2030 detectors, but could be detectable (albeit with a very low bit rate) after a few hundred years of GW detector development.

At the moment, the GW detector network is moving towards a curated open-data policy. The time series of the GW strain will be released from each data taking run with a cadence of ~1 year. The data rate from a single GW detector is ~1 TB/year (~5 TB/year for the year 2030 worldwide network). No groups have proposed to do SETI type searches on the GW data at this time. There are several groups working within the mainstream GW astrophysics community looking for long duration (t ~1 - 10,000 s), unmodeled signals from naturally occurring astrophysical sources.

### 5.2.5.2 MULTIMESSENGER TECHNOSIGNATURE SEARCHES

Multimessenger astronomy explores common sources of EM, gravitational, weak force, and strong force carriers in the forms of photons, gravitational waves, neutrinos, and cosmic rays (CRs). Although none of the facilities currently or soon to be involved in multimessenger astronomy were designed with technosignature searches in mind, they provide an opportunity for commensal technosignature work at low cost.

Gravitational waves and MeV or >TeV neutrinos are the most penetrating long-lived (hence propagating to astronomical distances) signatures of violent physical activity, and so might trace extremely high-energy events by technology, for instance the exploitation of naked singularities or evaporating black holes for experimental or practical purposes. The coincident detection of such carriers, which is already sought by programs such as AMON (the Astrophysical Multimessenger Observatory Network), may be useful to search for technosignatures alongside searches for naturally energetic events.

### 5.2.6 Emerging Technologies Relevant to SETI

The following emerging technologies may be relevant to technosignature searches in the solar system and beyond:

- Artificial intelligence,
- Laser propulsion and other advanced propulsion technologies,
- Data mining and storage technologies,
- Accelerated computations (GPUs, etc.),
- New detectors,
- Lunar far-side radio astronomy,
- Mid-frequency aperture arrays,
- Low-noise receivers, and
- Use of quantized photon angular momentum for communication.

In addition to new technologies to detect technosignatures, new algorithms are needed to accelerate technosignature searches in current and future data. Here, machine learning, Bayesian neural networks, inversions, etc. are proven to be useful. Also, advances in data archiving, curation, and storage (warehouse) are needed for high-efficiency data analysis.

## 5.3 WAYS TO CATALYZE SIGNIFICANT ADVANCES

In this section, various aspects of possible partnerships of NASA with private, public (including other federal agencies), philanthropic, and academic organizations are addressed. These partnerships are seen as a way of amplifying NASA investment into the technosignature field as well as to assure a possibility for achieving stability in funding technosignature research.

Also, development of post-detection information infrastructure is important. For example, one may use services and components (mostly commercially available) to support large-scale citizen science after a detection of a signal with content. Millions of users would want to access the data and may contribute significantly to efforts in analysis and comprehension.

## 5.3.1 Private Partnerships

In times of lack of federal funding for technosignature research, several private organizations have declared and pursued their interests to accelerate the technosignature field. The SETI Institute,[5] Breakthrough Foundation,[6] and PLANETS Foundation[7] have, among others, developed novel techniques, technologies, and sponsorships with the goal to detect technosignature and biosignature on nearby exoplanets. Industry is developing impactful commercial software tools in the area of big data, machine learning, anomaly detection, and deep learning. These computational techniques are highly relevant to the scientific community, which, in turn, provides interesting use-cases back to the commercial sector resulting in partnerships. An example of this is the partnership between exoplanet scientists and Google Brain developers that resulted in open source code to process Kepler data as well as new exoplanet discoveries.[8,9] Commercial partnerships are fostered by incubators like the Frontier Development Lab[10] hosted by NASA and the SETI Institute. NASA partnerships with the private sector can accelerate technology developments as well as their scientific and commercial applications. They can also provide valuable training for young researchers and technology entrepreneurs.

Another opportunistic pathway may be the creation of a prestigious competition for a lucrative monetary prize. The XPrize is an extraordinary example of what can be achieved with private funding and a global goal. Often the fiscal investment by multiple teams is far greater than the privately funded value of the prize.

## 5.3.2 Philanthropic and Academic Partnerships

The technosignature field has been decimated by the cessation of public funding. Expertise needs to be rebuilt via career development. Here, partnerships with, and opportunities for, the academic sector would help to create a pipeline of talented researchers for the future.

Funding for fellowships, internships, coops, and curriculum development at the national level can be achieved via NASA partnerships with NSF, the National Academies of Science, Engineering, and Medicine (NAS), and other federal agencies. Other sources include private foundations with a national or regional scope, state/local level university foundations, other local foundations, and individual donors. In addition, crowdfunding can be employed at both national and state/local levels.

### 5.3.2.1 TECHNOSIGNATURE REFERENCE LIBRARY INITIATIVE

An accessible, reliable, long-term database for technosignature work that has been done is needed. The database could be similar to the NASA Astrophysics Data System (ADS) effort, but should take into account the fact that technosignature research is a highly interdisciplinary field. It should include refereed papers, conference proceedings, and various relevant reports (like this one) on technosignaures. A recent introduction to interdisciplinary work related to SETI/technosignatures was compiled by Oman-Reagan et al. (2018). A searchable database of all published searches for technosignatures will be made public in January 2019 at https://technosearch.seti.org/.

---

[5] https://www.seti.org/

[6] https://breakthroughinitiatives.org/

[7] https://www.planets.life/

[8] https://ai.googleblog.com/2018/03/open-sourcing-hunt-for-exoplanets.html

[9] https://www.blog.google/technology/ai/hunting-planets-machine-learning/

[10] http://www.frontierdevelopmentlab.org

#### 5.3.2.2 Library of Congress Blumberg Chair in Astrobiology Program

This program, co-sponsored by NASA and the Library of Congress, fosters "research at the intersection of the science of astrobiology and its humanistic and societal implications," which can overlap into technosignatures. Chair-holders occupy the position for one year or less, however, and the research and events produced do not necessarily filter deeply into the science community. There may be ways to amplify the extent and impact of this program for both biosignatures and technosignatures.

#### 5.3.2.3 MIT "The Center for Brains, Minds and Machines" (CBMM)

This is an NSF-funded Science and Technology Center focused on an interdisciplinary study of intelligence. This effort is a multi-institutional collaboration headquartered at the McGovern Institute for Brain Research at MIT, with managing partners at Harvard University. It is focused specifically on modelling human intelligence but nonetheless relevant to modeling intelligence in the context of technosignature searches.

#### 5.3.2.4 Social Dimensions and Implications of the Research

Partnerships with academic institutions with strong humanities and social sciences representation are essential to grow and sustain the interdisciplinary expertise needed to explore the ongoing societal impact of technosignature searches (successful or not) and related questions. There are a number of U.S. universities with the necessary academic profile.

Philanthropic support has historically funded some interdisciplinary work related to SETI/technosignatures, and in the future, there may be foundations and donors that would be interested in partnering with a NASA-supported technosignature and biosignature initiatives to support interdisciplinary researchers and students.

It would be wise to consider the public not just as the passive recipients of the "impact" of detections, but as an active force with opinions that should be considered. A recent significant public consultation process was undertaken by NASA in collaboration with researchers from Arizona State University's Consortium for Science, Policy, and Outcomes (Tomblin et al. 2017). This would be a substantial undertaking but it could be a useful consultation process for designing a NASA-funded technosignature program, which truly takes into account current public interests and concerns (instead of presuming to already know what those are). Existing technical and humanistic expertise related to the search for technosignatures should be involved, as well as experts in this consultation method.

### 5.3.3 Partnerships with Other Federal Programs

Partnerships with other federal programs could foster a *sustained* interdisciplinary work, which should be integral to the search for technosignatures, concerning the following issues:

- Existing and potential societal impacts of the search for life in the universe (Dick, ed. 2015);
- Evolutionary phenomena that generate intelligence and social and technological complexity; and
- Computational linguistics and models of communication.

Each of these fields can inform and accelerate the search but will also be acutely relevant in the event that something is found.

#### 5.3.3.1 NSF Directorate for the Social, Behavioral and Economic Sciences (NSF/SBE) and National Endowment for the Humanities

Collaboration with these programs could promote a continually self-renewing ethical culture for all the fields embedded within the search for life (including, but not limited to, technosignatures). Of particular interest in this regard are the SBE Office of Multidisciplinary Activities programs in Cultivating Ethical Cultures for STEM (Science, Technology, Engineering and Mathematics), and the Science of Science & Innovation Policy (SciSIP) program. There is also an opportunity for integration with the NSF Scientific Integrity / Media Policy. Scientists engaged in the search for technosignatures have often been operating outside of frameworks of human ethics or bioethics, simply because astronomy is not generally subject to review or restrictions in this regard. (SETI has traditionally subscribed more to the metalaw framework.) However, as the subfields of technosignatures and biosignatures evolve and merge, and as the chances of detecting life grow, ethics are increasingly relevant. The technosignature research community would benefit from an engagement with scientific ethics frameworks from fields outside of astronomy.

Specialist research into public reactions and best communication practices regarding candidate signals and detections will significantly inform the future of technosignatures, and this research should be undertaken professionally within national human participants research ethics frameworks.

#### 5.3.3.2 NSF Directorate for Biological Sciences (NSF/BIO), and NSF Geosciences (NSF/GEO), including, but not limited to, the BIO/GEO Division of Emerging Frontiers (e.g., Rules of Life), NSF/BIO Evolutionary Processes Cluster, AccelNet

*Sustained* partnerships with these programs could facilitate the much-needed integration of current bioscience-based theory, method, and data into the physics-oriented technosignature field. The field will benefit enormously from effective interdisciplinary bridges. The search for technosignatures is predicated on the assumption that evolutionary processes have produced intelligent, complex, technology-making life elsewhere in the universe. The astronomy-based search for technosignatures has not required and does not necessarily require a detailed or current understanding of those evolutionary processes—in the same way that one can find a fossil or an ancient artifact on Earth without a comprehensive knowledge of vertebrate evolution or human history. However, deeper understandings of “what life does” will certainly inform and refine search strategies (particularly if signs of life continue to be difficult to find), and are *essential* for the interpretation of any significant findings (i.e., ‘who’ created the technosignature and what can be known or inferred about them). Interpretations will be critical both for science and for national and global public interest. Finally, technosignature and biosignature searches have effectively begun merging due to escalations/shifts in remote sensing capabilities; an integrated body of theory and an interconnected, cross-trained research community will facilitate progress in the years and decades to come.

#### 5.3.3.3 NSF Office of International Science and Engineering (NSF/OISE)

OISE-managed initiatives, such as the program “Accelerating Research through International Network-to-Network Collaborations” (AccelNet), may be relevant to the search for technosignatures.

#### 5.3.3.4 NSF-wide Programs

Other, general NSF programs aiming at broadening the impact of science on the society may be relevant to technosignature research, for example, the program “Increasing the Participation and Advancement of Women in Academic Science and Engineering Careers” (ADVANCE).

#### 5.3.3.5 U.S. Department of Commerce: NOAA and NIST

There could be opportunities for NASA to amplify its technosignature research via partnerships with other Federal agencies. Through such partnerships, relevant scientific and commercialization assets of other Federal agencies could complement assets used by NASA to conduct technosignature research. For example, two agencies that NASA could partner with could be the National Oceanic and Atmospheric Administration (NOAA) and the National Institute of Standards and Technology (NIST), both of which are under the U.S. Department of Commerce. Below are a few examples of such partnerships.

Partnering with the NOAA:

- The archives of comprehensive oceanic, atmospheric, and geophysical data of the NOAA’s National Centers for Environmental Information (NOAA 2018a) could be utilized in a complementary manner to inform theoretical and applied research conducted by NASA into technosignatures, whether in context of the Earth’s immediate environs or that of Earth-like exoplanets. Coincidentally, the NOAA and NASA have an existing partnership through NASA’s Joint Agency Satellite Division (NASA 2018).
- Federal agency Small Business Innovation Research (SBIR) programs provide training and technical assistance to small businesses to help them identify feasible businesses opportunities in terms of federal contracting in conjunction with commercial applications; this while assessing and mitigating business risks. Such programs could also provide enhanced opportunities for certified minority- and women-owned small businesses to participate. The SBIR program of the NOAA's Technology Partnerships Office (NOAA 2018b) could provide small business referrals to the NASA SBIR/STTR (Small Business Technology Transfer) program, where the referrals demonstrate a likelihood of advancing technosignature research and its commercial applications. Whether involving SBIR or similar activities, the advancement of technosignature research and its commercial applications could occur in areas such as communications and navigation; sensors, detectors and instruments; advanced telescope technologies; information technology for science data; and so on.

Partnering with the NIST:

- The measurement science and commercialization expertise of the NIST Technology Partnership Office (NIST 2018) could be utilized in a complementary manner to inform NASA's technosignature research, similar to the aforementioned example involving a NASA-NOAA partnership.
- The NIST SBIR program could provide referrals to the NASA SBIR/STTR program, similar to the example involving a NASA-NOAA partnership.

#### 5.3.3.6 SMITHSONIAN NATIONAL AIR AND SPACE MUSEUM

Philanthropic support and public outreach for NASA technosignature research could be accomplished in part through a partnership with the Smithsonian National Air and Space Museum. Such a partnership could entail Museum staff and fellows providing research support (via the Museum's Center for Earth and Planetary Sciences), designated gifts from Museum donors, and a public multimedia exhibit illustrating the history of NASA's efforts to search for life elsewhere in the universe (e.g., from the first ground-based radio telescopes to modern-day space satellites, etc.). Such an exhibit could be a prototype that other institutions could publicly showcase in support of technosignature research.

#### 5.3.3.7 DATA IN NASA ARCHIVES OF RELEVANCE TO TECHNOSIGNATURE SCIENCE

NASA has collected large data archives that may be relevant to technosignature searches, including asteroid searches (near-Earth objects) and planetary radar data. Anomaly detection in archival data is supported by programs, such as Astrophysics Data Analysis Program (ADAP). Post-detection dissemination of Information is relevant to Defense Advanced Research Projects Agency (DARPA).

# 6 References

## Section 1 References

Dyson, Freeman J. 1960. “Search for Artificial Stellar Sources of Infrared Radiation.” *Science* 131 (3414):1667–1668. doi: 10.1126/science.131.3414.1667.

Tarter, Jill. 2001. “The search for extraterrestrial intelligence (SETI).” *Annual Review of Astronomy and Astrophysics* 39 (1):511–548.

Tarter, Jill C. 2007. “The evolution of life in the Universe: are we alone?” *Highlights of Astronomy* 14:14–29.

## Section 2 References

Abeysekara, AU, S Archambault, Avery Archer, W Benbow, Ralph Bird, M Buchovecky, JH Buckley, Karen Byrum, Joshua V Cardenzana, and M Cerruti. 2016. “A search for brief optical flashes associated with the SETI target KIC 8462852.” *The Astrophysical Journal Letters* 818 (2):L33.

Annis, J, S Kent, F Castander, D Eisenstein, J Gunn, R Kim, R Lupton, R Nichol, M Postman, and W Voges. 1999. “The maxBCG technique for finding galaxy clusters in SDSS data.” Bulletin of the American Astronomical Society.

Arnold, Luc FA. 2005. “Transit light-curve signatures of artificial objects.” *The Astrophysical Journal* 627 (1):534.

Backus, Peter R. 1998. “The Phoenix search results at Parkes.” *Acta astronautica* 42 (10-12):651–654.

Bannister, Keith W, Tara Murphy, Bryan M Gaensler, and John E Reynolds. 2012. “Limits on prompt, dispersed radio pulses from gamma-ray bursts.” *The Astrophysical Journal* 757 (1):38.

Carrigan Jr, Richard A. 2009. “IRAS-based whole-sky upper limit on Dyson spheres.” *The Astrophysical Journal* 698 (2):2075.

Chyba, Christopher F, and Kevin P Hand. 2005. “Astrobiology: the study of the living universe.” *Annu. Rev. Astron. Astrophys.* 43:31–74.

Davies, PCW, and RV Wagner. 2013. “Searching for alien artifacts on the moon.” *Acta Astronautica* 89:261–265.

Drake, Frank, John H Wolfe, and Charles L Seeger. 1984. “SETI Science Working Group Report.”

Dreher, John W, and D Kent Cullers. 1997. “SETI figure of merit.” IAU Colloq. 161: Astronomical and Biochemical Origins and the Search for Life in the Universe.

Dyson, Freeman J. 1960. “Search for Artificial Stellar Sources of Infrared Radiation.” *Science* 131 (3414):1667–1668. doi: 10.1126/science.131.3414.1667.

Enriquez, J Emilio, Andrew Siemion, Griffin Foster, Vishal Gajjar, Greg Hellbourg, Jack Hickish, Howard Isaacson, Danny C Price, Steve Croft, and David DeBoer. 2017. “The breakthrough listen search for intelligent life: 1.1–1.9 GHz observations of 692 nearby stars.” *The Astrophysical Journal* 849 (2):104.

Forgan, Duncan H. 2013. “On the possibility of detecting class A stellar engines using exoplanet transit curves.” *arXiv preprint arXiv:1306.1672*.

Gray, Robert H, and Kunal Mooley. 2017. “A VLA Search for Radio Signals from M31 and M33.” *The Astronomical Journal* 153 (3):110.

Griffith, Roger L, Jason T Wright, Jessica Maldonado, Matthew S Povich, Steinn Sigurđsson, and Brendan Mullan. 2015. “The Ĝ infrared search for extraterrestrial civilizations with large energy supplies. III. The reddest extended sources in WISE.” T*he Astrophysical Journal Supplement Series* 217 (2):25.

Haqq-Misra, Jacob, and Ravi Kumar Kopparapu. 2012. “On the likelihood of non-terrestrial artifacts in the Solar System.” *Acta Astronautica* 72:15–20.

Harp, GR, Jon Richards, Seth Shostak, JC Tarter, Douglas A Vakoch, and Chris Munson. 2016. “Radio SETI observations of the anomalous star KIC 8462852.” *The Astrophysical Journal* 825 (2):155.

Howard, Andrew W, Paul Horowitz, David T Wilkinson, Charles M Coldwell, Edward J Groth, Norm Jarosik, David W Latham, Robert P Stefanik, Alexander J Willman Jr, and Jonathan Wolff. 2004. “Search for nanosecond optical pulses from nearby solar-type stars.” *The Astrophysical Journal* 613 (2):1270.

Howard, Andrew, Paul Horowitz, Curtis Mead, Pratheev Sreetharan, Jason Gallicchio, Steve Howard, Charles Coldwell, Joe Zajac, and Alan Sliski. 2007. “Initial results from Harvard all-sky optical SETI.” *Acta Astronautica* 61 (1-6):78–87.

Jugaku, Jun, and Shiro Nishimura. 1991. “A search for Dyson spheres around late-type stars in the IRAS catalog.” In *Bioastronomy The Search for Extraterrestial Life—The Exploration Broadens*, 295–298. Springer.

Korpela, Eric J, Shauna M Sallmen, and Diana Leystra Greene. 2015. “Modeling Indications of Technology in Planetary Transit Light Curves—Dark-Side Illumination.” *The Astrophysical Journal* 809 (2):139.

Lazio, J. 2008. “Radio wavelength transients: Current and emerging prospects.” *Astronomische Nachrichten: Astronomical Notes* 329 (3):330–333.

Margot, Jean-Luc, Adam H Greenberg, Pavlo Pinchuk, Akshay Shinde, Yashaswi Alladi, Srinivas Prasad, M Oliver Bowman, Callum Fisher, Szilard Gyalay, and Willow McKibbin. 2018. “A Search for Technosignatures from 14 Planetary Systems in the Kepler Field with the Green Bank Telescope at 1.15–1.73 GHz.” *The Astronomical Journal* 155 (5):209.

Mead, Curtis Charles. 2013. A Configurable Terasample-persecond Imaging System for Optical SETI. Doctoral dissertation, Harvard University.

Schmidt, Gavin A, and Adam Frank. 2018. “The Silurian hypothesis: would it be possible to detect an industrial civilization in the geological record?” *International Journal of Astrobiology*:1–9.

Schuetz, Marlin, Douglas A Vakoch, Seth Shostak, and Jon Richards. 2016. “Optical SETI observations of the anomalous star KIC 8462852.” *The Astrophysical Journal Letters* 825 (1):L5.

Schuetz, Marlin. 2018. “Recent Developments at the Boquete Optical SETI Observatory and Owl Observatory.” *arXiv preprint arXiv:1809.01956*.

Seeger, Charles L, and John H Wolfe. 1985. “Seti: The Microwave Search Problem and the Targeted Search Approach.” Symposium-International Astronomical Union.

Tarter, Jill. 2001. “The search for extraterrestrial intelligence (SETI).” *Annual Review of Astronomy and Astrophysics* 39 (1):511–548.

Tarter, J. C., A. Agrawal, R. Ackermann, P. Backus, S. K. Blair, M. T. Bradford, G. R. Harp, J. Jordan, T. Kilsdonk, K. E. Smolek, J. Richards, J. Ross, G. S. Shostak, and D. Vakoch. 2010. “SETI turns 50: five decades of progress in the search for extraterrestrial intelligence.” SPIE Optical Engineering + Applications.

Tellis, Nathaniel K, and Geoffrey W Marcy. 2015. “A search for optical laser emission using Keck HIRES.” *Publications of the Astronomical Society of the Pacific* 127 (952):540.

Tellis, Nathaniel K, and Geoffrey W Marcy. 2017. “A search for laser emission with megawatt thresholds from 5600 FGKM stars.” *The Astronomical Journal* 153 (6):251.

Tingay, SJ, C Tremblay, Andrew Walsh, and Ryan Urquhart. 2016. “An Opportunistic Search for ExtraTerrestrial Intelligence (SETI) with the Murchison Widefield Array.” *The Astrophysical Journal Letters* 827 (2):L22.

Tingay, SJ, CD Tremblay, and S Croft. 2018. “A Search for Extraterrestrial Intelligence (SETI) toward the Galactic Anticenter with the Murchison Widefield Array.” *The Astrophysical Journal* 856 (1):31.

Von Korff, J, P Demorest, E Heien, E Korpela, D Werthimer, J Cobb, M Lebofsky, D Anderson, B Bankay, and A Siemion. 2013. “Astropulse: A Search for Microsecond Transient Radio Signals Using Distributed Computing. I. Methodology.” *The Astrophysical Journal* 767 (1):40.

Wolfe, John H, J Billingham, RE Edelson, RB Crow, S Gulkis, ET Olsen, BM Oliver, and AM Peterson. 1981. “SETI-the search for extraterrestrial intelligence-plans and rationale.” NASA Conference Publication.

Wright, Shelley A, Frank Drake, Remington PS Stone, Dick Treffers, and Dan Werthimer. 2001. “Improved optical SETI detector.” The Search for Extraterrestrial Intelligence (SETI) in the Optical Spectrum III.

Wright, Shelley A, Dan Werthimer, Richard R Treffers, Jérôme Maire, Geoffrey W Marcy, Remington PS Stone, Frank Drake, Elliot Meyer, Patrick Dorval, and Andrew Siemion. 2014a. “A near-infrared SETI experiment: instrument overview.” Ground-based and Airborne Instrumentation for Astronomy V.

Wright, J. T., R. L. Griffith, S. Sigurdsson, M. S. Povich, and B. Mullan. 2014b. “The Ĝ Infrared Search for Extraterrestrial Civilizations with Large Energy Supplies. II. Framework, Strategy, and First Result.” *The Astrophysical Journal* 792 (1):27.

Wright, Jason T, Kimberly Cartier, Ming Zhao, Daniel Jontof-Hutter, and Eric B Ford. 2016. “The Search for Extraterrestrial Civilizations with Large Energy Supplies. IV. The Signatures and Information Content of Transiting Megastructures.” *The Astrophysical Journal* 816.

Wright, Jason T, Shubham Kanodia, and Emily G Lubar. 2018. “How Much SETI Has Been Done? Finding Needles in the n-Dimensional Cosmic Haystack.” *arXiv preprint arXiv:1809.07252*.

Zackrisson, Erik, Per Calissendorff, Saghar Asadi, and Anders Nyholm. 2015. “Extragalactic seti: The tully–fisher relation as a probe of dysonian astroengineering in disk galaxies.” *The Astrophysical Journal* 810 (1):23.

## Section 3 References

Carrigan Jr, Richard A. 2009. “IRAS-based whole-sky upper limit on Dyson spheres.” *The Astrophysical Journal* 698 (2):2075.

Gray, Robert H, and Kunal Mooley. 2017. “A VLA Search for Radio Signals from M31 and M33.” *The Astronomical Journal* 153 (3):110.

Lubin, Philip 2016. “The Search for Directed Intelligence.” *Reviews in Human Space Exploration* 1:20-45

Lubin, Philip, et al. 2016. “Implications of Directed Energy for SETI.” *Proceedings of the SPIE* 9981:99810H

Schwartz, R. N., and Charles H Townes. 1961. “Interstellar and interplanetary communication by optical masers.” *Nature* 190 (4772):205.

Stewart, A. et al. 2017. “The Trillion Planet Survey.” *Proceedings of the SPIE* 10401:104010C

Tingay, SJ, C Tremblay, Andrew Walsh, and Ryan Urquhart. 2016. “An Opportunistic Search for ExtraTerrestrial Intelligence (SETI) with the Murchison Widefield Array.” *The Astrophysical Journal Letters* 827 (2):L22.

Tingay, SJ, CD Tremblay, and S Croft. 2018. “A Search for Extraterrestrial Intelligence (SETI) toward the Galactic Anticenter with the Murchison Widefield Array.” *The Astrophysical Journal* 856 (1):31.

Villarroel, Beatriz, Inigo Imaz, and Josefine Bergstedt. 2016. “Our Sky Now and Then: Searches for Lost Stars and Impossible Effects as Probes of Advanced Extraterrestrial Civilizations.” *The Astronomical Journal* 152 (3):76.

Walkowicz, Lucianne, AR Howe, R Nayar, EL Turner, J Scargle, V Meadows, and A Zee. 2014. “Mining the Kepler Data using Machine Learning.” American Astronomical Society Meeting Abstracts # 223.

## Section 4 References

Bialy, Shmuel, and Abraham Loeb. 2018. “Could Solar Radiation Pressure Explain `Oumuamua’s Peculiar Acceleration?” *The Astrophysical Journal Letters* 868 (1):L1.

Blount, Zachary D, Richard E Lenski, and Jonathan B Losos. 2018. “Contingency and determinism in evolution: Replaying life’s tape.” *Science* 362 (6415):eaam5979.

Bolin, Bryce T, Harold A Weaver, Yanga R Fernandez, Carey M Lisse, Daniela Huppenkothen, R Lynne Jones, Mario Jurić, Joachim Moeyens, Charles A Schambeau, and Colin T Slater. 2017. “APO time-resolved color photometry of highly elongated interstellar object 1I/`Oumuamua.” *The Astrophysical Journal Letters* 852 (1):L2.

Boumghar, Redouane, Zahi Kakish, Drew Bischel, Ana Mosquera, Francisco Rodríguez Lera, Phil Metzger, J-L Galache, Sylvester Kaczmarek, and Tim Seabrook. 2018. “NASA Frontier

Development Lab Technical Memorandum for Space Resources Prospection: Cooperative Multi-Agent Systems for Mission Planning."

Carlotto, Mark J. 1988. "Digital imagery analysis of unusual Martian surface features." *Applied Optics* 27 (10):1926-1933.

Carlotto, Mark J. 1997. "Evidence in support of the hypothesis that certain objects on Mars are artificial in origin." *Journal of Scientific Exploration* 11 (2):1–26.

Clayton, N. R. Powell, C. Logan and I. Mikhalevich, eds. 2017. Theme issue, "'Convergent minds: the evolution of cognitive complexity in nature" *Interface Focus* 2017 vol 7 iss 3.

Davies, PCW, and RV Wagner. 2013. "Searching for alien artifacts on the moon." *Acta Astronautica* 89:261–265.

Denisenko, D, and V Lipunov. 2013. "MASDB2 identified with the man-made object." *The Astronomer's Telegram* 5616.b

Denning, Kathryn. 2009. "Social Evolution State of the field." Cosmos & Culture: Cultural Evolution in a Cosmic Context.

Enriquez, J Emilio, Andrew Siemion, Griffin Foster, Vishal Gajjar, Greg Hellbourg, Jack Hickish, Howard Isaacson, Danny C Price, Steve Croft, and David DeBoer. 2017. "The breakthrough listen search for intelligent life: 1.1–1.9 GHz observations of 692 nearby stars." *The Astrophysical Journal* 849 (2):104.

Enriquez, J Emilio, Andrew Siemion, T Joseph W Lazio, Matt Lebofsky, David HE MacMahon, Ryan S Park, Steve Croft, David DeBoer, Nectaria Gizani, and Vishal Gajjar. 2018. "Breakthrough Listen Observations of 1I/`Oumuamua with the GBT." *arXiv preprint arXiv:1801.02814*.

Freitas Jr, Robert A, and Francisco Valdes. 1980. "A search for natural or artificial objects located at the Earth-Moon libration points." *Icarus* 42 (3):442-447.

Freitas Jr, Robert A. 1983. "If they are here, where are they? Observational and search considerations." *Icarus* 55 (2):337–343.

Ginsburg, Idan, Manasvi Lingam, and Abraham Loeb. 2018. "Galactic Panspermia." *arXiv preprint arXiv:1810.04307*.

Haqq-Misra, Jacob, and Ravi Kumar Kopparapu. 2012. "On the likelihood of non-terrestrial artifacts in the Solar System." *Acta Astronautica* 72:15–20.

Harp, Gerry R, Jon Richards, Peter Jenniskens, Seth Shostak, and Jill C Tarter. 2018. "Radio SETI observations of the interstellar object `OUMUAMUA." *Acta Astronautica*.

Marino, Lori, and Kathryn Denning. 2012. "Virtual Workshops on Intelligence in Astrobiology." Two workshops, Arizona State University and Georgia Tech, in conjunction with Arizona State University Astrobiology team, NASA Astrobiology Institute.

Marino, Lori. 2015. "The landscape of intelligence." In *The Impact of Discovering Life Beyond Earth*, S. Dick, ed. Cambridge University Press pp 95–112.

Meech, Karen J, Robert Weryk, Marco Micheli, Jan T Kleyna, Olivier R Hainaut, Robert Jedicke, Richard J Wainscoat, Kenneth C Chambers, Jacqueline V Keane, and Andreea Petric. 2017. "A brief visit from a red and extremely elongated interstellar asteroid." *Nature* 552 (7685):378.

Morris, Simon Conway. 2011. “Predicting what extra-terrestrials will be like: and preparing for the worst.” *Philosophical Transactions of the Royal Society of London A: Mathematical, Physical and Engineering Sciences* 369 (1936):555–571.

Motesharrei, Safa, Jorge Rivas, and Eugenia Kalnay. 2014. “Human and nature dynamics (HANDY): Modeling inequality and use of resources in the collapse or sustainability of societies.” *Ecological Economics* 101:90–102.

Schmidt, Gavin A, and Adam Frank. 2018. “The Silurian hypothesis: would it be possible to detect an industrial civilization in the geological record?” *International Journal of Astrobiology*:1–9.

Trilling, David E, Tyler Robinson, Alissa Roegge, Colin Orion Chandler, Nathan Smith, Mark Loeffler, Chad Trujillo, Samuel Navarro-Meza, and Lori M Glaspie. 2017. “Implications for planetary system formation from interstellar object 1I/2017 U1 (`Oumuamua).” *The Astrophysical Journal Letters* 850 (2):L38.

Wright, Jason T. 2018. “Prior indigenous technological species.” *International Journal of Astrobiology* 17 (1):96–100.

## Section 5 References

Avin, Shahar, Bonnie C. Wintle, Julius Weitzdörfer, Seán S. Ó hÉigeartaigh, William J. Sutherland, and Martin J. Rees. 2018. “Classifying global catastrophic risks.” *Futures* 102:20-26. doi: https://doi.org/10.1016/j.futures.2018.02.001.

Balbi, Amedeo. 2018. “The Impact of the Temporal Distribution of Communicating Civilizations on Their Detectability.” *Astrobiology* 18 (1):54–58.

Berdyugina, SV, AV Berdyugin, DM Fluri, and V Piirola. 2011. “Polarized reflected light from the exoplanet HD189733b: first multicolor observations and confirmation of detection.” *The Astrophysical Journal Letters* 728 (1):L6.

Berdyugina, SV, JR Kuhn, M Langlois, G Moretto, J Krissansen-Totton, D Catling, JL Grenfell, Tina Santl-Temkiv, Kai Finster, and J Tarter. 2018a. “The Exo-Life Finder (ELF) telescope: New strategies for direct detection of exoplanet biosignatures and technosignatures.” Ground-based and Airborne Telescopes VII.

Berdyugina, Svetlana V, and Jeff R Kuhn. 2017. “Surface Imaging of Proxima b and Other Exoplanets: Topography, Biosignatures, and Artificial Mega-Structures.” *arXiv preprint arXiv:1711.00185*.

Berdyugina, Svetlana V, Andrei V Berdyugin, Dominique M Fluri, and Vilppu Piirola. 2007. “First detection of polarized scattered light from an exoplanetary atmosphere.” *The Astrophysical Journal Letters* 673 (1):L83.

Berdyugina, Svetlana V, Jeff R Kuhn, David M Harrington, Tina Šantl-Temkiv, and E John Messersmith. 2016. “Remote sensing of life: polarimetric signatures of photosynthetic pigments as sensitive biomarkers.” *International Journal of Astrobiology* 15 (1):45–56.

Berdyugina, Svetlana V, Jeff R Kuhn, Ruslan Belikov, and Slava G Turyshev. 2018b. “Exoplanet Terra Incognita.” *arXiv preprint arXiv:1809.05031*.

Berea, Anamaria. 2018. “An Atlas of Communication Evolution Based on a Unified Database of Evidence.” White paper for Decoding Alien Intelligence workshop, SETI Institute, March 2018 https://daiworkshop.seti.org/programagenda.

Berns, Gregory. 2018. “The Brain Ark: Brain Connectome Properties of Terrestrial Intelligence as a Metric for Extraterrestrial Intelligence.” White paper for Decoding Alien Intelligence workshop, SETI Institute, March 2018 https://daiworkshop.seti.org/programagenda.

Cocconi, Giuseppe, and Philip Morrison. 1959. “Searching for interstellar communications.” *Nature* 184 (4690).

Covault, Corbin E. 2001. “Large area solar power heliostat array for OSETI.” The Search for Extraterrestrial Intelligence (SETI) in the Optical Spectrum III.

De la Torre, Gabriel, and Manuel A Garcia. 2018. “The cosmic gorilla effect or the problem of undetected non terrestrial intelligent signals.” *Acta Astronautica* 146:83–91.

Deacon, Terrence. 2018. “Semiotic Universals: Beyond The Radical Translation Problem.” White paper for Decoding Alien Intelligence workshop, SETI Institute, March 2018 https://daiworkshop.seti.org/programagenda.

Denning, Kathryn. 2009. “Social Evolution State of the field.” Cosmos & Culture: Cultural Evolution in a Cosmic Context.

Denning, Kathryn. 2018. “ How Humans Matter Now: The Relevance of Anthropology and Archaeology for The New SETI.” White paper for Decoding Alien Intelligence workshop, SETI Institute, March 2018 https://daiworkshop.seti.org/programagenda.

Dick, Steven J. 2015. *The impact of discovering life beyond earth*: Cambridge University Press.

Doyle, Laurance R, Brenda McCowan, Simon Johnston, and Sean F Hanser. 2011. “Information theory, animal communication, and the search for extraterrestrial intelligence.” *Acta Astronautica* 68 (3–4):406-417.

Drake, Frank, John H Wolfe, and Charles L Seeger. 1984. “SETI Science Working Group Report.”

Dyson, Freeman J. 1960. “Search for Artificial Stellar Sources of Infrared Radiation.” *Science* 131 (3414):1667–1668. doi: 10.1126/science.131.3414.1667.

Elliott, John R, and Stephen Baxter. 2012. “The DISC quotient.” *Acta Astronautica* 78:20–25.

Elliott, John R. 2008. “A Post-Detection Decipherment Strategy.” 57th International Astronautical Congress, Glasgow, UK.

Elliott, John R. 2009. “A semantic ‘engine’for universal translation.” *Acta Astronautica*.

Elliott, John. 2011. “The filtration of inter-galactic objets trouvés and the identification of the lingua ex machina hierarchy.” *Acta Astronautica* 68 (3–4):399–405.

Elliott, John. 2012. “Constructing the matrix.” *Acta Astronautica* 78:26-30.

Elliott, John. 2015. “Beyond an anthropomorphic template.” *Acta Astronautica* 116:403–407.

Fluri, DM, and SV Berdyugina. 2010. “Orbital parameters of extrasolar planets derived from polarimetry.” *Astronomy & Astrophysics* 512:A59.

Forgan, Duncan H, and Martin Elvis. 2011. “Extrasolar asteroid mining as forensic evidence for extraterrestrial intelligence.” *International Journal of Astrobiology* 10 (4):307–313.

Frank, Adam, Axel Kleidon, and Marina Alberti. 2017. “Earth as a hybrid planet: The anthropocene in an evolutionary astrobiological context.” Anthropocene 19:13–21.

Grinspoon, David. 2016. *Earth in Human Hands: Shaping Our Planet's Future*: Grand Central Publishing.

Haqq-Misra, Jacob. 2014. “Damping of glacial-interglacial cycles from anthropogenic forcing.” *Journal of Advances in Modeling Earth Systems* 6 (3):950–955. doi: doi:10.1002/2014MS000326.

Herzing, Denise, Kathryn Denning, and John Elliott. 2018. “Beyond Initial Signal Categorization and Syntactic Assignment: the Role of Metadata in Decoding and Interpreting Nonhuman Signals.” 42nd COSPAR Scientific Assembly.

Kershenbaum, A, T Bergman, N Carlson, F da Cunha, L Doyle, J Elliott, R Ferrer-i-Cancho, M Gustison, G Harp, and B McCowan. “What Animal Studies Can Tell Us About Detecting Intelligent Messages From Outside Earth.” White paper for Decoding Alien Intelligence workshop, SETI Institute, March 2018 https://daiworkshop.seti.org/programagenda.

Kipping, David M, and Alex Teachey. 2016. “A cloaking device for transiting planets.” *Monthly Notices of the Royal Astronomical Society* 459 (2):1233–1241.

Kissell, Kenneth E. 1974. “Polarization effects in the observation of artificial satellites.” IAU Colloq. 23: Planets, Stars, and Nebulae: Studied with Photopolarimetry.

Krissansen-Totton, Joshua, Stephanie Olson, and David C Catling. 2018. “Disequilibrium biosignatures over Earth history and implications for detecting exoplanet life.” *Science advances* 4 (1):eaao5747.

Kuhn, Jeff R, and Svetlana V Berdyugina. 2015. “Global warming as a detectable thermodynamic marker of Earth-like extrasolar civilizations: the case for a telescope like Colossus.” *International journal of astrobiology* 14 (3):401–410.

Kuhn, JR, G Moretto, SV Berdyugina, M Langlois, and E Thiebaut. 2018. “The Exo-Life Finder (ELF) Telescope: Design and Synthetic Beam Concepts.” SPIE Astronomical Telescopes+ Instrumentation.

Langlois, Maud, Arthur Vigan, Kjetil Dohlen, Claire Moutou, Jean-Luc Beuzit, A Boccaletti, Michael Carle, Anne Costille, R Dorn, and Laurence Gluck. 2014. “Infrared Differential Imager and Spectrograph for SPHERE: Performance assessment for on-sky operation.” Ground-based and Airborne Instrumentation for Astronomy V.

Lin, Henry W, Gonzalo Gonzalez Abad, and Abraham Loeb. 2014. “Detecting industrial pollution in the atmospheres of earth-like exoplanets.” *The Astrophysical Journal Letters* 792 (1):L7.

Lineweaver, Charles H, Yeshe Fenner, and Brad K Gibson. 2004. “The galactic habitable zone and the age distribution of complex life in the Milky Way.” *Science* 303 (5654):59–62.

Lingam, Manasvi, and Abraham Loeb. 2017. “Natural and artificial spectral edges in exoplanets.” *Monthly Notices of the Royal Astronomical Society*: Letters.

Loeb, Abraham, and Edwin L Turner. 2012. “Detection technique for artificially illuminated objects in the outer solar system and beyond.” *Astrobiology* 12 (4):290–294.

Loeb, Abraham, Rafael A Batista, and David Sloan. 2016. “Relative likelihood for life as a function of cosmic time.” *Journal of Cosmology and Astroparticle Physics* 2016 (08):040.

Marino, Lori. 2015. *The landscape of intelligence*: Cambridge University Press Cambridge.

McShea, Daniel. 2018. “Universals in the Evolution of Intelligence.” White paper for Decoding Alien Intelligence workshop, SETI Institute, March 2018 https://daiworkshop.seti.org/programagenda.

Moretto, G, JR Kuhn, JF Capsal, D Audigier, M Langlois, K Thetpraphi, SV Berdyugina, and D Halliday. 2018. “The Exo-Life Finder (ELF) Telescope: Optical Concept and Hybrid Dynamic Live-Optical Surfaces.” Proc. SPIE.

NASA. 2018. Joint Agency Satellite Division. Retrieved from https://science.nasa.gov/about-us/smd-programs/joint-agency-satellite-division

NIST. 2018. About the Technology Partnerships Office. Retrieved from https://www.nist.gov/tpo/about-technology-partnerships-office

NOAA. 2018a. About the National Centers for Environmental Information. Retrieved from https://www.ncei.noaa.gov/about

NOAA. 2018b. Small Business Innovation Research program. Retrieved from https://techpartnerships.noaa.gov/SBIR

Oman-Reagan, M. P., Denning, K., Fraknoi, A., Tarter, J. C., Traphagan, J. W., Lempert, W. 2018. SETI Primer: Interdisciplinary Introduction to SETI. Google document. https://docs.google.com/document/d/1Vd5llZrXOjKZSKZi9-_LWRERvKO4fQh59zPiRcXkJ3s/edit#

Rees, Martin J. 2003. *Our final hour: A scientist's warning: How terror, error, and environmental disaster threaten humankind's future in this century--on Earth and beyond*: Basic Books (AZ).

Rose, Christopher, and Gregory Wright. 2004. “Inscribed matter as an energy-efficient means of communication with an extraterrestrial civilization.” *Nature* 431:47. doi: 10.1038/nature02884.

Schneider, Jean, Alain Léger, Malcolm Fridlund, Glenn J White, Carlos Eiroa, Thomas Henning, Tom Herbst, Helmut Lammer, René Liseau, and Francesco Paresce. 2010. “The far future of exoplanet direct characterization.” *Astrobiology* 10 (1):121–126.

Schwartz, RN, and Charles H Townes. 1961. “Interstellar and interplanetary communication by optical masers.” *Nature* 190 (4772):205.

Shannon, Claude Elwood. 1948. “A mathematical theory of communication.” *Bell system technical journal* 27 (3):379–423.

Snik, Frans, Julia Craven-Jones, Michael Escuti, Silvano Fineschi, David Harrington, Antonello De Martino, Dimitri Mawet, Jérôme Riedi, and J Scott Tyo. 2014. “An overview of polarimetric sensing techniques and technology with applications to different research fields.” Polarization: Measurement, Analysis, and Remote Sensing XI.

Stevens, Adam, Duncan Forgan, and Jack O'Malley James. 2016. “Observational signatures of self-destructive civilizations.” *International Journal of Astrobiology* 15 (4):333–344.

Tarter, J. C., A. Agrawal, R. Ackermann, P. Backus, S. K. Blair, M. T. Bradford, G. R. Harp, J. Jordan, T. Kilsdonk, K. E. Smolek, J. Richards, J. Ross, G. S. Shostak, and D. Vakoch. 2010. “SETI turns 50: five decades of progress in the search for extraterrestrial intelligence." SPIE Optical Engineering + Applications.

Tomblin, David, Zachary Pirtle, Mahmud Farooque, David Sittenfeld, Erin Mahoney, Rick Worthington, Gretchen Gano, Michele Gates, Ira Bennett, and Jason Kessler. 2017. "Integrating Public Deliberation into Engineering Systems: Participatory Technology Assessment of NASA's Asteroid Redirect Mission." *Astropolitics* 15 (2):141–166.

Turchin, Peter, Thomas E Currie, Harvey Whitehouse, Pieter François, Kevin Feeney, Daniel Mullins, Daniel Hoyer, Christina Collins, Stephanie Grohmann, and Patrick Savage. 2018. "Quantitative historical analysis uncovers a single dimension of complexity that structures global variation in human social organization." *Proceedings of the National Academy of Sciences* 115 (2):E144–E151.

Wilson, Edward O. 2002. *The future of life*: Vintage.

Zhang, Yunfan Gerry, Vishal Gajjar, Griffin Foster, Andrew Siemion, James Cordes, Casey Law, and Yu Wang. 2018. "Fast Radio Burst 121102 Pulse Detection and Periodicity: A Machine Learning Approach." *The Astrophysical Journal* 866 (2):149.

# 7 ACRONYMS

| | |
|---|---|
| 2MASS | Two Micron All-Sky Survey |
| AAC | artificial atmospheric constituent |
| AccelNet | Accelerating Research through International Network-to-Network Collaborations |
| ADAP | Astrophysics Data Analysis Program |
| ADS | NASA Astrophysics Data System |
| AI | artificial intelligence |
| ALMA | Atacama Large Millimeter/submillimeter Array |
| AMON | Astrophysical Multimessenger Observatory Network |
| AO | Arecibo Observatory |
| ARAA | Annual Review of Astronomy and Astrophysics |
| ARIEL | Atmospheric Remote-sensing Infrared Exoplanet Large-survey |
| ASKAP | Australian SKA Pathfinder |
| ATA | Allen Telescope Array |
| BIO | NSF Directorate for Biological Sciences |
| BOINC | Berkeley Open Infrastructure for Network Computing |
| CASPER | Collaboration for Radio Astronomy Signal Processing and Electronics Research |
| CBMM | MIT Center for Brains, Minds and Machines |
| CFC | chlorofluorocarbon |
| CR | cosmic ray |
| DARPA | Defense Advanced Research Projects Agency |
| DFM | Drake figure of merit |
| EIRP | equivalent isotropic radiated power |
| ELF | Exo-Life Finder |
| ELT | Extremely Large Telescope |
| FFT | fast Fourier transform |
| FOM | figure of merit |
| FOV | field of view |
| FPGA | field programmable gate array |
| FRB | fast radio burst |
| GBT | Green Bank Telescope |
| GEO | NSF Geosciences |
| GMRT | Giant Metrewave Radio Telescope |
| GO | guest observer |
| GPU | graphics processing unit |
| GUPPI | Green Bank Ultimate Pulsar Processing Instrument |
| GW | Gravitational Wave |
| HabEx | Habitable Exoplanet Observatory |
| HARPS | High Accuracy Radial velocity Planet Searcher |
| HERA | Hydrogen Epoch of Reionization Array |
| HIRES | High Resolution Echelle Spectrometer |
| HPF | Habitable Zone Planet Finder |
| HST | Hubble Space Telescope |
| IRAS | Infrared Astronomical Satellite |
| JWST | James Webb Space Telescope |
| LiDAR | Light Detection and Ranging |
| LIGO | Laser Interferometer Gravitational-Wave Observatory |
| LMT | Large Millimetre Telescope |
| LOFAR | Low-Frequency Array |
| LoFASM | Low Frequency All Sky Monitor |
| LRS | Low Resolution Spectrometer |
| LSST | Large Synoptic Survey Telescope |
| LUVOIR | Large UV/Optical/IR Surveyor |
| METI | Messaging Extraterrestrial Intelligence |
| MIT | Massachusetts Institute of Technology |
| MOST | Microvariability and Oscillations of STars telescope |
| MWA | Murchison Wide-field Array |
| NAI | NASA Astrobiology Institute |
| NAS | National Academies of Science, Engineering, and Medicine |
| NASA | National Aeronautics and Space Administration |
| NEID | NN-explore Exoplanet Investigations with Doppler spectroscopy |
| NEOWISE | Near-Earth Object Wide-field Survey Explorer |
| ngVLA | next-generation Very Large Array |
| NIR | near-infrared |
| NIROSETI | Near-Infrared Optical SETI |
| NIST | National Institute of Standards and Technology |
| NN-EXPLORE | NASA-NSF Exoplanet Observational Research |
| NOAA | National Oceanic and Atmospheric Administration |
| NSF | National Science Foundation |
| OIR | Optical/InfraRed |
| OISE | NSF Office of International Science and Engineering |
| Pan-STARRS | Panoramic Survey Telescope and Rapid Response System |
| POP | persistent organic pollutant |
| PRV | precision radial velocity |
| RFI | radio frequency interference |
| SBE | NSF Directorate for the Social, Behavioral and Economic Sciences |
| SBIR | Small Business Innovation Research |
| SciSIP | Science of Science & Innovation Policy |
| SERENDIP | Search for Extraterrestrial Radio Emissions from Nearby Developed Intelligent Populations |
| SETI | Search for Extraterrestrial Intelligence |
| SKA | Square Kilometre Array |
| SNR | signal-to-noise ratio |
| STEM | Science, Technology, Engineering and Mathematics |
| STTR | NASA SBIR Small Business Technology Transfer |
| UAV | unmanned aerial vehicle |
| UCLA | University of California, Los Angeles |
| UCO | University of California Observatories |
| UVOIR | ultraviolet, optical, infrared |
| VASCO | Vanishing & Appearing Sources during a Century of Observations |
| VERITAS | Very Energetic Radiation Imaging Telescope Array System |
| VLT | Very Large Telescope |
| WISE | Wide-field Infrared Survey Explorer |
| ZTF | Zwicky Transient Facility |